\documentclass[journal, twocolumn, a4paper]{IEEEtran}

\usepackage{cite}
\usepackage{url}
\usepackage{graphicx}
\usepackage{amssymb,amsmath,bm}
\usepackage{textcomp}
\usepackage{booktabs}
\usepackage{hyperref}
\usepackage{verbatim}
\usepackage{multirow}
\usepackage[table,xcdraw]{xcolor}
\usepackage{array}
\usepackage{tabularx}
\usepackage{color} 
\newcolumntype{Y}{>{\centering\arraybackslash}X}
\newcommand{\norm}[1]{\left\lVert#1\right\rVert}

\begin{document}

\title{Quasi-Periodic Parallel WaveGAN: A Non-autoregressive Raw Waveform Generative Model with Pitch-dependent Dilated Convolution Neural Network}

\author{Yi-Chiao Wu,
        Tomoki Hayashi,
        Takuma Okamoto,~\IEEEmembership{Member,~IEEE,}
        Hisashi Kawai,~\IEEEmembership{Member,~IEEE,}\\
        and~Tomoki Toda,~\IEEEmembership{Member,~IEEE}
\thanks{Manuscript received xxx xx, 2020; revised xxx xx, 2020. This work was supported in part by the Japan Science and Technology Agency (JST), Precursory Research for Embryonic Science and Technology (PRESTO) under Grant JPMJPR1657, in part by the JST, CREST under Grant JPMJCR19A3, and in part by the Japan Society for the Promotion of Science (JSPS) Grants-in-Aid for Scientific Research (KAKENHI) under Grant 17H06101. The initial investigation in this study was performed while Y.-C. Wu was interning at NICT.}
\thanks{Y.-C. Wu is with Graduate School of Informatics, Nagoya University, Aichi, Japan (e-mail: yichiao.wu@g.sp.m.is.nagoya-u.ac.jp).}
\thanks{T. Hayashi is with Graduate School of Information Science, Nagoya University, Aichi, Japan (e-mail: hayashi.tomoki@g.sp.m.is.nagoya-u.ac.jp).}
\thanks{T. Okamoto and H. Kawai are with National Institute of Information and Communications Technology, kyoto, Japan (e-mail: okamoto@nict.go.jp, hisashi.kawai@nict.go.jp).}
\thanks{T. Toda is with Information Technology Center, Nagoya University, Aichi, Japan (e-mail: tomoki@icts.nagoya-u.ac.jp).}
}

\markboth{Journal of \LaTeX\ Class Files,~Vol.~0, No.~0, July~2020}%
{Shell \MakeLowercase{\textit{et al.}}: Bare Demo of IEEEtran.cls for IEEE Journals}

\maketitle

\begin{abstract}
In this paper, we propose a quasi-periodic parallel WaveGAN (QPPWG) waveform generative model, which applies a quasi-periodic (QP) structure to a parallel WaveGAN (PWG) model using pitch-dependent dilated convolution networks (PDCNNs). PWG is a small-footprint GAN-based raw waveform generative model, whose generation time is much faster than real time because of its compact model and non-autoregressive (non-AR) and non-causal mechanisms. Although PWG achieves high-fidelity speech generation, the generic and simple network architecture lacks pitch controllability for an unseen auxiliary fundamental frequency ($F_{0}$) feature such as a scaled $F_{0}$. To improve the pitch controllability and speech modeling capability, we apply a QP structure with PDCNNs to PWG, which introduces pitch information to the network by dynamically changing the network architecture corresponding to the auxiliary $F_{0}$ feature. Both objective and subjective experimental results show that QPPWG outperforms PWG when the auxiliary $F_{0}$ feature is scaled. Moreover, analyses of the intermediate outputs of QPPWG also show better tractability and interpretability of QPPWG, which respectively models spectral and excitation-like signals using the cascaded fixed and adaptive blocks of the QP structure.
\end{abstract}

\begin{IEEEkeywords}
Neural vocoder, parallel WaveGAN, quasi-periodic WaveNet, pitch-dependent dilated convolution
\end{IEEEkeywords}

\section{Introduction}

\IEEEPARstart{S}{peech} generation is a technique to generate specific speech according to given inputs such as texts (text-to-speech, TTS), the speech of a source speaker (speaker voice conversion, VC), and noisy speech (speech enhancement, SE). The core of speech generation is the controllability of speech components, and the fundamental technique is called a vocoder~\cite{vocoder_1939, vocoder_1966, phase_vocoder}. A vocoder encodes speech into acoustic representations such as spectral and prosodic features and then decodes specific speech on the basis of the manipulated acoustic features. Conventional vocoders such as STRAIGHT~\cite{straight} and WORLD~\cite{world} are based on a source-filter model~\cite{source_filter}, which models speech with vocal fold movements (excitation) and vocal tract resonances (spectral envelope). However, many oversimplified designs such as a fixed length of the analysis window, a time-invariant linear filter, and a stationary Gaussian process are imposed on the conventional vocoders. The losses of phase information and temporal details caused by these ad~hoc designs result in speech quality degradation.

To tackle these problems, many neural network (NN)-based speech generation models~\cite{wavenet, samplernn, fftnet, wavernn, lpcnet, pwn, clarinet, waveglow, flowavenet, waveffjord, glotgan_2017, glotgan_2019, gelp, nhv, hooligan, hinet, pwngan, sa_pwngan, pwg, melgan, gantts, vocgan, multi-melgan,  pap_gan, nsf_2019, nsf_2020, qpnet_2019, qpnet_2020} have been proposed. In contrast to the conventional source-filter-based vocoders, most of these models directly model the relationships among speech waveform samples. Specifically, autoregressive (AR) models such as WaveNet (WN)~\cite{wavenet} and SampleRNN~\cite{samplernn} achieve high-fidelity speech generation by modeling the probability distribution of each speech sample with the given auxiliary features and previous samples. Taking conventional-vocoder-extracted acoustic features as the auxiliary features for NN-based speech generation models~\cite{sd_wn_vocoder, si_wn_vocoder, ns_wn_vocoder, sp_wn_vocoder, srnn_vocoder}, which replace the synthesizer of the conventional vocoders, also achieved early success. However, the AR mechanism and huge network architectures of WN and SampleRNN result in very slow generations, making these models impractical for realistic scenarios. To tackle these problems, many compact AR models with specific knowledge~\cite{fftnet, wavernn, lpcnet} and non-AR models such as flow-based~\cite{pwn, clarinet, waveglow, flowavenet, waveffjord} and generative adversarial network (GAN)-based~\cite{pwngan, pwg, melgan, vocgan, multi-melgan, gantts, glotgan_2017, glotgan_2019, gelp, sa_pwngan, hinet, nhv, hooligan} models have been proposed.

\begin{figure*}[ht]
\centering
\centerline{\includegraphics[width=2\columnwidth]{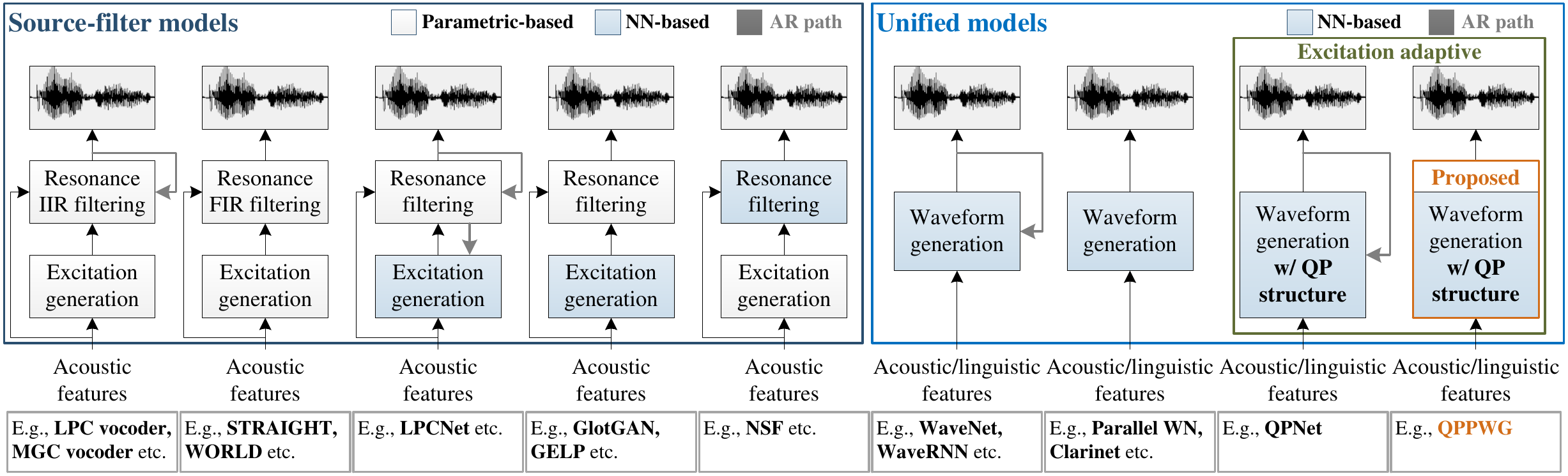}}
\caption{Comparison of waveform generative architectures.}
\label{fig:vocoders}
\end{figure*}

Although these NN-based models achieve high-fidelity speech generation without many ad~hoc designs, the data-driven nature, the generic network architecture, and the lack of prior acoustic knowledge of these models make most of them lose acoustic controllability and robustness to unseen auxiliary features~\cite{vc_wn_2017, nu_p_2018, nu_np_2018, cl_2018, cl_2020}. For instance, without explicitly modeling the excitation signals as conventional source-filter models, it is difficult for WN to generate speech with accurate pitches outside the fundamental frequency ($F_{0}$) range of training data when conditioned on the scaled $F_{0}$ feature~\cite{qpnet_2019, qpnet_2020}. However, using carefully designed mixed periodic and aperiodic inputs and source-filter-like architectures, the authors of~\cite{pap_gan, nsf_2019, nsf_2020} proposed different NN-based models attaining pitch controllability. In our previous works~\cite{qpnet_2019, qpnet_2020}, we also proposed a quasi-periodic WN (QPNet), which has a conventional-vocoding-like framework while using a unified network without the requirement of specific mixed inputs. QPNet advances the dilated convolution neural networks (DCNNs)~\cite{dcnn} of WN with a pitch-dependent mechanism to improve the pitch controllability of WN by dynamically changing the network architecture according to the auxiliary $F_{0}$ feature.

Although QPNet markedly improves the pitch accuracy of the generated speech, the AR mechanism and the huge network requirement of WN result in slow generations. To address this problem, we extend the AR PDCNN of QPNet to a non-AR PDCNN and apply the quasi-periodic (QP) structure to parallel WaveGAN (PWG)~\cite{pwg}, which is a compact non-AR model with a WN-like network architecture consisting of stacked DCNN layers. The proposed QPPWG speech generation model~\cite{qppwg_2020_1} attains pitch controllability using a simple pitch-dependent architecture without the requirement of specific mixed periodic and aperiodic inputs as in~\cite{pap_gan, nsf_2019, nsf_2020}. Although QPPWG greatly improved the pitch controllability of PWG, the behind mechanisms of the QP structure in the non-AR model, the characteristic of each component, and the effective architectures are remained to be explored.

Therefore, in this paper, we conduct more evaluations with several hyperparameter settings and network architectures to comprehensively explore the efficiency of model structures and the internal behaviors and mechanisms of QPPWG. Specifically, model details such as the order of the cascaded structure, the numbers of dilation cycles and residual blocks, and the balanced ratio of adaptive and fixed modules are investigated. Both objective and subjective evaluations are conducted, and the experimental results show the effectiveness of the proposed QP structure for PWG. Furthermore, we also investigate a new parallel QP structure and show the reason why the stacked QP structure is selected for QPPWG. In addition, comprehensive analyses of intermediate outputs of QPPWG are presented to make us know more about the internal behaviors of the QP network. The discussions of the QP structure understanding show the tractability and interpretability of QPPWG. The analyses confirm our assumption that QPPWG respectively models harmonic components with long-term correlations and non-harmonic components with short-term correlations using the adaptive module with pitch-dependent DCNNs (PDCNNs) and the fixed module with DCNNs of QPPWG.

This paper is organized as follows. In Section~\ref{related}, we review the recent GAN-based neural vocoders. In Section~\ref{pwg}, a brief introduction to PWG is presented. In Section~\ref{qppwg}, we describe the concepts and details of the proposed QPPWG. In Section~\ref{experiment}, objective and subjective tests are presented to show the effectiveness of QPPWG for generating speech with scaled $F_{0}$. Further discussion of QPPWG is presented in Section~\ref{discussion}. Finally, the conclusion is given in Section~\ref{conclusion}.

\section{Related Work} \label{related}

\subsection{Source-filter and Data-driven Vocoders}
Because of the high temporal resolution of speech signals, directly modeling raw speech waveforms is challenging. One of the standard speech modeling methods is source-filter modeling~\cite{source_filter}. Specifically, the speech generative process is formulated as a convolution of an excitation (voice source) signal and a spectral filter. The excitation signal models the glottal waveform generated by vocal fold movements, and the spectral filter models vocal tract resonances. As shown in Fig.~\ref{fig:vocoders}, the conventional parametric vocoders generate speech samples in an AR manner such as LPC vocoders~\cite{lpc_1982, lpc_1995} and mel-generalized cepstrum (MGC) vocoders~\cite{mgc_1983, mgc_1992} or in a non-AR manner such as STRAIGHT~\cite{straight} and WORLD~\cite{world}. Motivated by the development of deep NNs, NN-based excitation generation models with the AR mechanism such as LPCNet~\cite{lpcnet} and the non-AR mechanism such as GlotGAN~\cite{glotgan_2017, glotgan_2019} and GELP~\cite{gelp} have been proposed to improve the generated speech quality. Moreover, the authors of~\cite{nsf_2019} and ~\cite{nsf_2020} also proposed a neural source-filter (NSF) network to model the source-filter generative framework with an advanced neural filter.

In addition to the source-filter-based vocoders, many unified NN-based waveform generative models have been proposed to directly generate high-fidelity speech waveforms from acoustic features in a purely data-driven manner as shown in Fig.~\ref{fig:vocoders}. For example, the WN~\cite{wavenet} and WaveRNN~\cite{wavernn} models autoregressively generate speech samples conditioned on acoustic/linguistic features and the previous samples, and the non-AR Parallel WN~\cite{pwn} and Clarinet~\cite{clarinet} models simultaneously generate all speech samples with acoustic/linguistic features and white noise inputs. Although these models achieve high-fidelity speech generation without many ad~hoc designs imposed on them, pitch controllability is degraded because of the data-driven nature of not explicitly modeling excitation signals as the source-filter-based models. To improve the pitch controllability while keeping the unified and generic network architectures, we proposed a QP structure~\cite{qpnet_2019, qpnet_2020} for WN. The proposed QPNet implemented a source-filter-like mechanism into WN to simultaneously model the periodicity and aperiodicity of speech signals using a pitch-adaptive network architecture. In this paper, to achieve real-time generations, we extend the QP structure to the non-AR PWG model~\cite{pwg} to markedly improve the generation speed and show the generality of the proposed PDCNN, which can be easily integrated into any CNN-based network.

\subsection{GAN-based Vocoders}
Recently, because of the successes of GAN~\cite{gan} in image and video generation, GAN-based neural vocoders~\cite{pap_gan, pwg, melgan, vocgan, multi-melgan, gantts, glotgan_2017, glotgan_2019, gelp, pwngan, sa_pwngan, hinet, hooligan, nhv} have also been proposed. The two main categories of recent GAN-based neural vocoders are models with prior speech knowledge and models directly trained in a data-driven manner as mentioned in the previous section.

Among the models with speech knowledge, GlotGAN~\cite{glotgan_2017, glotgan_2019} achieved early success in generating glottal excitation signals, but it suffered severe speech quality degradation when directly applied to raw speech waveform generation. GELP~\cite{gelp} has been proposed to improve the glottal generator by using short-time Fourier transform (STFT)-based regression loss and the adversarial loss of the final generated waveforms. For neural spectral filtering, the authors of~\cite{pap_gan} proposed a GAN-based vocoder with tailored periodic and aperiodic inputs, and the model was trained with the GAN loss of the generated waveform and the Gaussian loss of its aperiodic components. Inspired by the neural excitation generation of differentiable digital signal processing (DDSP)~\cite{ddsp} and the neural spectral filtering of NSF, completely differentiable source-filter vocoders with a GAN structure such as neural homomorphic vocoder (NHV)~\cite{nhv} and HooliGAN~\cite{hooligan} also have been proposed. Furthermore, the authors of HiNet~\cite{hinet} also adopt a deep NN (DNN) model and an NSF model with GAN structures to respectively predict amplitude spectrum and phase for hierarchical speech generation.

Among the purely data-driven models, teacher--student-based parallel WN~\cite{pwn} conditioned on the mel-spectrogram has been combined with a GAN structure of the waveform domain for joint optimization~\cite{pwngan} and speaker adaptation~\cite{sa_pwngan}. Furthermore, MelGAN~\cite{melgan} and GAN-TTS~\cite{gantts} have been proposed to directly transform acoustic features to speech waveforms using GAN structures with tailored generators and discriminators. Specifically, both MelGAN and GAN-TTS have an upsampling generator that gradually expands the temporal resolution of the input acoustic features to match the speech waveforms. MelGAN adopts a multi-scale discriminator with several different downsampling rates to enable its generator to capture the information of different levels. GAN-TTS also adopts an ensemble of 10 similar discriminators with different input window sizes with or without the conditional acoustic features to guide its generator to learn different aspects of speech information. Furthermore, the variants of MelGAN such as VocGAN~\cite{vocgan} adopted a multi-scale generator and a hierarchically-nested discriminator and multi-band MelGAN~\cite{multi-melgan} incorporated a multi-band technique into MelGAN also achieved further speech quality or generative efficiency improvements.

Another purely data-driven model called PWG~\cite{pwg}, which transforms white noise into speech with conditional mel-spectrograms, has also been proposed. Instead of complex discriminators, PWG adopts a simple one with stacked DCNN layers. To achieve stable PWG training, STFT-based losses are also utilized. In conclusion, most recent GAN-based neural vocoders have adopted a convolutional feedforward network, and the hierarchical information of speech waveforms such as multi-resolution STFT-based losses is essential for training a high-quality raw waveform generator.

In this paper, we focus on introducing prior pitch knowledge to the data-driven PWG model, which is fast, compact, simple, and easy to train, to improve its pitch controllability and speech modeling capability and make it more consistent with the definition of a vocoder.

\begin{figure}[t]
\centering
\centerline{\includegraphics[width=0.77\columnwidth]{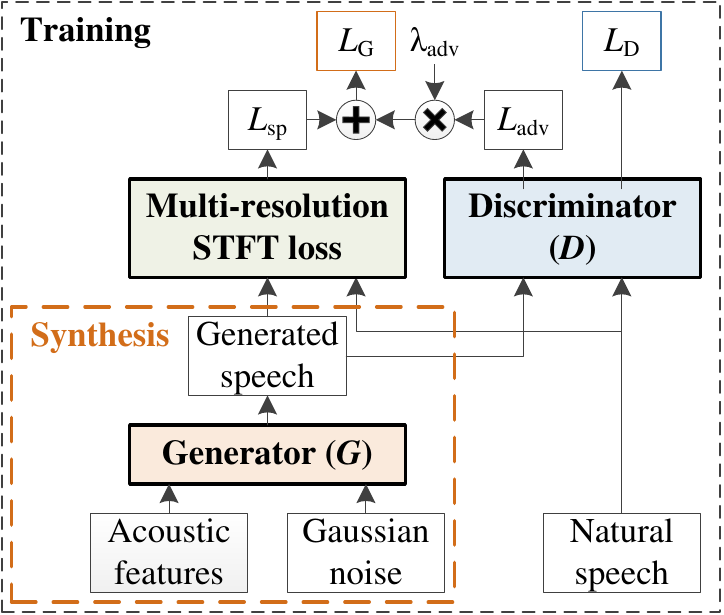}}
\caption{Architecture of parallel WaveGAN.}
\label{fig:pwg}
\end{figure}

\section{Parallel WaveGAN} \label{pwg}

As shown in Fig.~\ref{fig:pwg}, PWG includes a classical GAN module, which consists of a discriminator ($D$) and a generator ($G$), with fully convolutional feedforward networks and an additional multi-resolution STFT loss module. The details are as follows.

\subsection{GAN-based Waveform Generation}
A WN-like architecture is adopted for the generator of PWG. The main differences between the PWG generator and WN are a Gaussian noise input instead of previous samples, a raw waveform output instead of a probability distribution, and a non-AR manner. Specifically, the inputs of the generator are a Gaussian noise sequence $\boldsymbol{z}$ and auxiliary acoustic features, and $\boldsymbol{z}$ is drawn from a Gaussian distribution with zero mean and standard deviation, denoted as $N(0, I)$. The output of the generator is the waveform samples. The generator, which tries to generate realistic speech samples, is trained in a manner adversarial to the discriminator, which attempts to distinguish natural ($real$) and generated ($fake$) speech waveforms. The adversarial loss of the generator ($ L_{\mathrm{adv}}$) is formulated as
\begin{align}
L_{\mathrm{adv}}(G, D)=\mathbb{E}_{\boldsymbol{z} \in N(0, I)}\left[(1-D(G(\boldsymbol{z})))^{2}\right].
\label{eq:ladv}
\end{align}
Note that all auxiliary features of the generator are omitted in this section for simplicity. Unlike some flow-based models~\cite{waveglow, flowavenet}, which adopt an invertible network to map the real data into the Gaussian noise sequence, the generator of PWG learns to transfer the input noise sequence to the output waveforms via the feedback from the discriminator.

Furthermore, a simple architecture consisting of stacked DCNN layers with LeakyReLU~\cite{leakyrelu} activation functions is adopted for the discriminator of PWG, and the dilation size of each DCNN layer increases exponentially with a base of 2 and the exponent of its layer index. The discriminator is trained to minimize the adversarial loss ($ L_{\mathrm{D}}$) formulated as 
\begin{align}
&L_{\mathrm{D}}(G, D) \nonumber \\
&=\mathbb{E}_{\boldsymbol{x} \in p_{\mathrm{data}}}\left[(1-D(\boldsymbol{x}))^{2}\right]+\mathbb{E}_{\boldsymbol{z} \in N(0, I)}\left[D(G(\boldsymbol{z}))^{2}\right],
\label{eq:ld}
\end{align}
where $\boldsymbol{x}$ denotes the natural samples and $p_{\mathrm{data}}$ denotes the data distribution of the natural samples.

\subsection{Multi-resolution STFT Loss}
Since training PWG with only adversarial losses is difficult and tends to be unstable, an additional STFT-based loss ($ L_{\mathrm{sp}}$) is adopted to improve the stability and efficiency of the GAN training. Specifically, a spectral convergence loss ($ L_{\mathrm{sc}}$) is formulated as
\begin{align}
L_{\mathrm{sc}}(\boldsymbol{x}, \hat{\boldsymbol{x}})
&=\frac{\norm {|\mathrm{STFT}(\boldsymbol{x})|-|\mathrm{STFT}(\hat{\boldsymbol{x}})|}_{F}}
{\norm{|\mathrm{STFT}(\boldsymbol{x}) |}_{F}},
\label{eq:lsc}
\end{align}
and a log STFT magnitude loss ($ L_{\mathrm{mag}}$) is formulated as
\begin{align}
&L_{\mathrm{mag}}(\boldsymbol{x}, \hat{\boldsymbol{x}}) \nonumber \\
&=\frac{1}{N}\norm{\log |\mathrm{STFT}(\boldsymbol{x})|-\log
|\mathrm{STFT}(\hat{\boldsymbol{x}})|}_{L1},
\label{eq:lmag}
\end{align}
where $\hat{\boldsymbol{x}}$ denotes the samples generated from the generator, $\norm{\cdot}_{F}$ is the Frobenius norm, $\norm{\cdot}_{L1}$ is the $L1$ norm, $\left | \mathrm{STFT}\left ( \cdot  \right ) \right |$ denotes the STFT magnitudes, and ${N}$ is the number of magnitude elements. The multi-resolution STFT-based loss $ L_{\mathrm{sp}}$ is formulated as 
\begin{align}
L_{\mathrm{sp}}(G)=\frac{1}{M}\sum_{m=1}^{M}(L^{(m)}_{\mathrm{sc}}(G)+L^{(m)}_{\mathrm{mag}}(G)),
\label{eq:lsp}
\end{align}
where $M$ denotes the number of STFT setting groups, and each group includes different FFT sizes, frame lengths, and frame shifts. The losses $L^{(m)}_{\mathrm{sc}}$ and $L^{(m)}_{\mathrm{mag}}$ are calculated on the basis of the STFT features extracted using the settings of the $m$  group. The multiple STFT losses prevent the generator from a suboptimal problem and enhance the modeling capability of the generator by making it capture speech structures with different resolutions. In conclusion, the overall training loss of the PWG generator ($ L_{\mathrm{G}}$) is formulated as 
\begin{align}
L_{\mathrm{G}}(G, D)=L_{\mathrm{sp}}(G)+\lambda_{\mathrm{adv}} L_{\mathrm{adv}}(G, D),
\label{eq:lg}
\end{align}
which is a weighted sum of $ L_{\mathrm{adv}}$ and $ L_{\mathrm{sp}}$ with weight $\lambda_{\mathrm{adv}}$. The hyperparameter $\lambda_{\mathrm{adv}}$ is empirically set to 4.0 in this paper.

\subsection{Problems in Using PWG as a Vocoder}
Although PWG achieves high-fidelity speech generation with acoustic features, it is still vulnerable to unseen acoustic features such as scaled $F_{0}$. That is, the speech quality and pitch accuracy of the PWG-generated speech will markedly degrade when the $F_{0}$ of the auxiliary acoustic features is scaled or is outside the $F_0$ range of training data~\cite{qpnet_2019, qpnet_2020}. The possible reasons for the degradation are the generic architecture, data-driven nature, and lack of prior speech knowledge. Moreover, since speech is a quasi-periodic signal, which includes both periodic components with long-term correlations and aperiodic components with short-term correlations, modeling both components with the fixed network architecture of PWG is inefficient. For instance, the fixed {\it receptive field} size of the network for both periodic and aperiodic components may not be reasonable, and the {\it receptive field} may include many redundant samples when modeling the periodic structures of speech.

\section{Quasi-Periodic Parallel WaveGAN} \label{qppwg}

Since pitch controllability is an essential feature of a vocoder, we propose QPPWG~\cite{qppwg_2020_1} to improve the pitch controllability and speech modeling efficiency of PWG. Specifically, because the effectiveness of the GAN structure and the multi-resolution STFT losses have been shown for PWG, the proposed QPPWG only improves the generator of PWG using the QP structure while keeping other components of PWG the same. The QP structure of the proposed generator introduces pitch information to the network via a non-AR PDCNN module and a cascaded architecture. The details are as follows. 

\begin{figure}[t]
\centering
\centerline{\includegraphics[width=1.0\columnwidth]{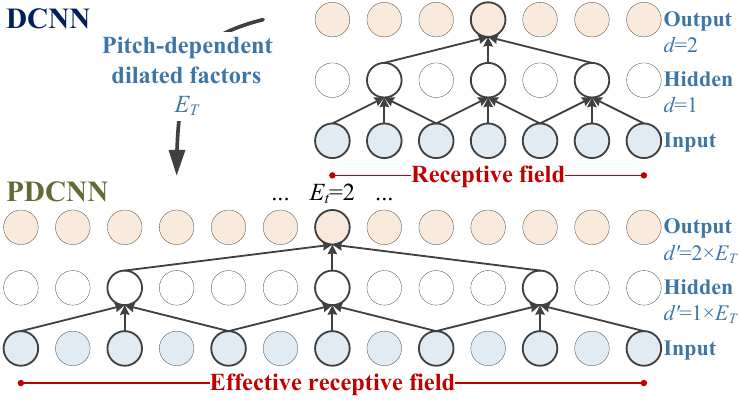}}
\caption{Pitch-dependent dilated convolution.}
\label{fig:pdcnn}
\end{figure}

\begin{figure*}[t]
\centering
\centerline{\includegraphics[width=2\columnwidth]{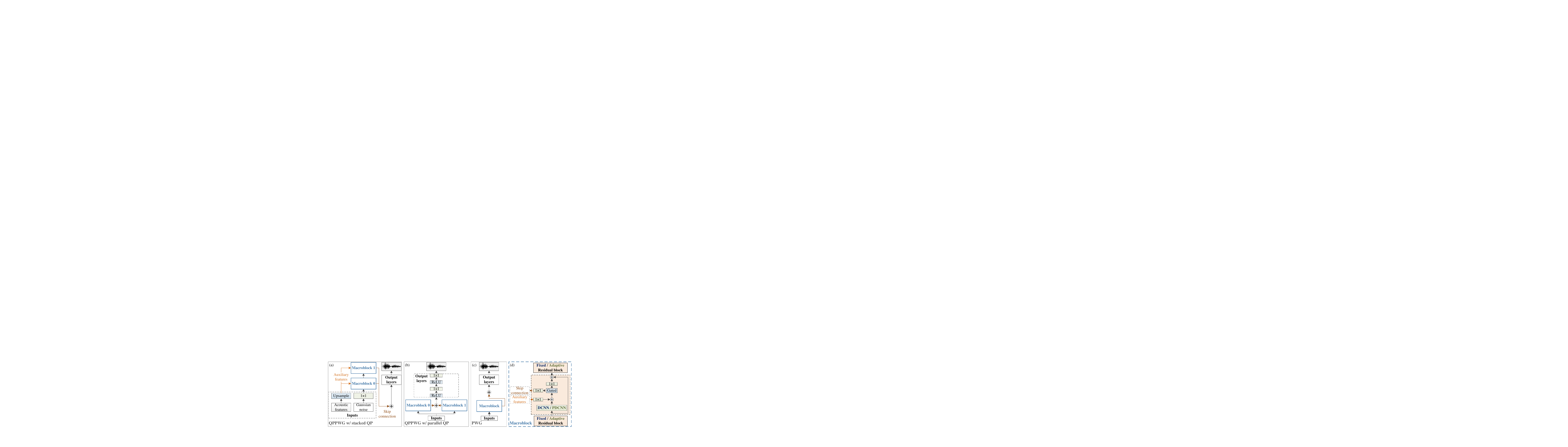}}
\caption{(a) QPPWG generator with stacked QP strucutre; (b) QPPWG generator with parallel QP strucutre; (c) PWG generator (d) Macroblock.}
\label{fig:qppwg}
\end{figure*}

\subsection{Non-autoregressive Pitch-dependent Dilated Convolution}
Inspired by pitch filtering in code-excited linear prediction (CELP)~\cite{celp_1984, celp_1985}, we proposed a PDCNN for causal AR models~\cite{qpnet_2019, qpnet_2020}. In this paper, we further extend the PDCNN in a non-causal manner for the non-AR PWG model. As shown in Fig.~\ref{fig:pdcnn}, a DCNN is a convolution layer with gaps between input samples, and the length of each gap is a predefined hyperparameter called the dilation size (rate). The non-causal dilated convolution can be formulated as
\begin{align}
 \boldsymbol{y}_{t}^{(\mathrm{o})}
=\boldsymbol{W}^{(\mathrm{c})}\times\boldsymbol{y}_{t}^{(\mathrm{i})}
+\boldsymbol{W}^{(\mathrm{p})}\times\boldsymbol{y}_{t-d}^{(\mathrm{i})}
+\boldsymbol{W}^{(\mathrm{f})}\times\boldsymbol{y}_{t+d}^{(\mathrm{i})},
\label{eq:dcnn}
\end{align}
where $\boldsymbol{y}_{t}^{(\mathrm{o})}$ is the DCNN output at sample $t$, $\boldsymbol{y}_{t}^{(\mathrm{i})}$ is the DCNN input at sample $t$, and $d$ is the dilation size. $\boldsymbol{W}^{(\mathrm{c})}$, $\boldsymbol{W}^{(\mathrm{p})}$, and $\boldsymbol{W}^{(\mathrm{f})}$ are the trainable {$1\times1$} convolution filters of the current, previous, and following samples, respectively. For the vanilla DCNN, $d$ is a predefined time-invariant constant. As an extension of a DCNN, the dilation size $d^{\prime}$ of a PDCNN is pitch-dependent and time-variant. 

Specifically, the pitch-dependent dilated factor $E_{t}$ is multiplied by the dilation size $d$ in each time step $t$ to dynamically set the dilation size $d^{\prime}$ as
\begin{align}
d^{\prime}=E_{t} \times d.
\label{eq:dp}
\end{align}
The dilated factor $E_{t}$ is derived from
\begin{align}
E_{t}=F_{s}/(F_{0,t}\times a),
\label{eq:et}
\end{align}
where $F_{s}$ is the sampling rate, $F_{0, t}$ is the fundamental frequency of the input sample at time step $t$, and $a$ is the {\it dense factor}. The {\it dense factor} $a$ is a hyperparameter that indicates the number of samples in one cycle taken as the inputs of a PDCNN. The higher the {\it dense factor}, the lower the sparsity of the PDCNN. Using the pitch-dependent dilation size, the architecture of QPPWG with PDCNNs is dynamically changed according to the input $F_{0}$ feature. 

Furthermore, according to our previous work~\cite{ qpnet_2019, qpnet_2020}, calculating $E_t$ using the interpolated $F_0$ values of the adjacent voiced segments achieves higher speech quality than directly setting $E_t$ to one for the unvoiced segments. Because our internal evaluation results of QPPWG also show the same tendency, all QPPWG models in this paper adopt the interpolated $F_0$ values for calculating the $E_t$ values of the unvoice segments. In conclusion, the adaptive architecture of QPPWG introduces pitch knowledge to the network to improve the pitch controllability, allows each sample to have a specific {\it receptive field} size, and efficiently extends the {\it receptive fields}.

\subsection{QPPWG Generator with PDCNNs}
As shown in Fig.~\ref{fig:qppwg}, a QPPWG/PWG generator is composed of input, macroblock, and output modules. The input module includes a Gaussian noise input with $1\times1$ CNN and upsampled acoustic features with the matched temporal resolution to the output waveform samples. As shown in Fig.~\ref{fig:qppwg} (d), a macroblock includes several stacked residual blocks. The inputs of each residual block are the residual connection output of the previous block and auxiliary features. The outputs of each residual block are the residual connection output for the next block input and the skip connection to the output module. The architecture of each residual block consists of a DCNN/PDCNN layer, a gate structure, and a residual connection. Last, the summation of the skip connections from all residual blocks is processed by two ReLU~\cite{relu} activations with $1\times1$ CNNs to directly output speech waveform samples.

The main difference between the QPPWG and PWG generators is the QP structure. Specifically, a QPPWG generator includes a fixed macroblock and an adaptive macroblock while a PWG generator includes only one fixed macroblock. The fixed macroblock consists of only fixed (residual) blocks with DCNN layers, and the adaptive macroblock consists of only adaptive (residual) blocks with PDCNN layers. Each fixed block adopts a DCNN with a fixed network architecture to model the aperiodic speech components such as spectral envelopes with short-term correlations. Each adaptive block adopts a PDCNN layer to model the periodic speech components such as excitation signals with long-term correlations, and the PDCNN layer makes the architecture of the block adaptive to auxiliary $F_{0}$ values.

As shown in Fig.~\ref{fig:qppwg} (a), unlike PWG consisting of residual blocks with only DCNNs, QPPWG adopts a cascaded architecture composed of two different macroblocks. The cascaded architecture simultaneously models both periodic and aperiodic speech components in an efficient manner by using prior pitch knowledge, which also improves its pitch controllability. The cascaded architecture with prior pitch knowledge is assumed to have better tractability and interpretability than the original PWG architecture since it models different speech components with related specific network structures. Furthermore, in this paper, since we assume that the fixed and adaptive macroblocks respectively focus on aperiodic and periodic components, we also explore a new parallel QP structure as shown in Fig.~\ref{fig:qppwg} (b) to better understand the internal speech production mechanisms.

\section{Experiments} \label{experiment}

\subsection{Experimental Settings}
All speech generation models in this paper were trained in a multi-speaker manner. The training corpus consisted of 2200 utterances of the “slt” and “bdl” speakers of the CMU-ARCTIC corpus~\cite{arctic} and 852 utterances of all speakers of the Voice Conversion Challenge 2018 (VCC2018) corpus~\cite{vcc2018}. The total size of the training corpus was around 3000 utterances and the data length was around 2.5 hours. The testing corpus was the SPOKE set of the VCC2018 corpus. The SPOKE set consists of two male and two female speakers, and each speaker has 35 testing utterances. The sampling rate of all speech data was set to 22,050 Hz, and the resolution of the speech data was 16-bit.

The auxiliary features of these speech generation models were composed of one-dimensional continuous $F_{0}$, one-dimensional unvoiced/voiced binary code ($U/V$), 35-dimensional mel-cepstrum ($mcep$), and two-dimensional coded aperiodicity ($codeap$) features. Specifically, the WORLD (WD)\footnote{https://github.com/JeremyCCHsu/Python-Wrapper-for-World-Vocoder} vocoder was adopted to extract one-dimensional $F_{0}$ and 513-dimensional spectral ($sp$) and aperiodicity ($ap$) features with a frameshift of 5~ms. $F_{0}$ was interpolated to the continuous $F_{0}$ and converted to $U/V$, $ap$ was coded into $codeap$, and $sp$ was parameterized into $mcep$. To simulate unseen data, the continuous $F_{0}$ was scaled by ratios of 0.5 and 2 while keeping the other features the same. Moreover, the dilated factor $E_{t}$ of QPPWG was empirically calculated on the basis of the continuous $F_{0}$ because of the higher speech quality~\cite{ qpnet_2019, qpnet_2020}.

All PWG-like models were trained with the RAdam optimizer~\cite{radam} ($\epsilon=10^{-6}$) with 400~k iterations. Specifically, the generators were trained with only multi-resolution STFT losses for the first 100~k iterations and then jointly trained with the discriminators for the following 300~k iterations. The multi-resolution STFT losses were calculated on the basis of three STFT setting groups including different FFT sizes (1024/2048/512), frame shifts (120/240/50), and frame lengths (600/1200/240). The balanced weight  $\lambda_{\mathrm{adv}}$ of $ L_{\mathrm{adv}}$ was set to 4.0. The generators’ learning rate was $10^{-4}$ and the discriminators’ learning rate was $5 \times 10^{-5}$. Both learning rates decayed by 50~\% every 200~k iterations. The minibatch size was six and the batch length was 25,520 samples. Furthermore, the baseline QPNet\footnote{https://github.com/bigpon/QPNet} model was trained with the Adam optimizer~\cite{adam} with 200K iterations. The learning rate of QPNet was $10^{-4}$ without decay, and the minibatch size was one with a batch length of 20,000 samples.

\subsection{Model Descriptions}
In this paper, several variants of PWG and QPPWG models and a baseline QPNet model were involved in the evaluations. To describe the different architecture of each model, several basic modules are introduced. Specifically, a macroblock module consisting of stacked residual blocks was adopted, and each macroblock was only composed of one type of residual block namely, adaptive blocks (B$_\mathrm{Ada}$) or fixed blocks (B$_\mathrm{Fix}$). The PWG models only consisted of one macroblock (Macro 0) with fixed blocks. The proposed QPPWG and baseline QPNet models were composed of two cascaded macroblocks (Macro 0 and 1) with different types of residual block. 

Taking vanilla PWG as an example, the architecture composed of 30 fixed blocks with three cycles (repeats) of exponentially increasing dilation size, and each cycle contained 10 fixed blocks. Therefore, the number of total blocks (Block Num) of vanilla PWG was 30, and the vanilla PWG architecture was 10 fixed blocks $\times$ 3 cycles denoted as B$_\mathrm{Fix}10\times3$. For the baseline QPNet, since Macro 0 consisted of 12 fixed blocks with 3 cycles (B$_\mathrm{Fix}4\times3$), and Macro 1 consisted of 4 adaptive blocks with 1 cycles (B$_\mathrm{Ada}4\times1$), the order of macroblock (Macro order) was denoted as B$_\mathrm{Fix}\rightarrow$ B$_\mathrm{Ada}$. The proposed QPPWG models followed the same naming conventions.  

Moreover, all PWG and QPPWG models had the same discriminator architecture, which consisted of 10 non-causal DCNN layers with 64 convolution channels, three kernels, and LeakyReLU ($\alpha=0.2$) activation functions. For each adaptive/fixed block of the QPPWG/PWG generator, a gated activation with tanh and sigmoid functions was adopted, and the number of CNN channels of residual and skip connections and auxiliary features was also 64. The QPNet structure followed that in our previous works~\cite{qpnet_2020}, and the number of CNN channels of residual connections and auxiliary features was 512 and that of skip connections was 256.

\subsection{Objective Evaluations}

As reported in this section, the quality of the vocoders was evaluated by the mel-cepstral distortion (MCD), root mean square error (RMSE) of log $F_{0}$, and $U/V$ decision error. These measurements were calculated using the auxiliary features and the acoustic features extracted from the generated speech. Specifically, WD disentangles speech into a resonance component, spectral envelope $sp$, and source components including $F_{0}$ and $ap$, and the designs of WD try to make the extracted spectral envelope and source components highly uncorrelated~\cite{cheaptrick}. Therefore, the $mcep$ and $F_{0}$ features were assumed as independent in this paper. In other words, for the auxiliary features of the neural vocoders, we only manipulated the $F_{0}$ values and kept other acoustic features the same. Since the topic of this paper is a neural vocoder, the ground truth acoustic features of the objective evaluations were the auxiliary features. That is, even for the scaled $F_{0}$ scenarios, the ground truth $mcep$ was still the $mcep$ extracted from natural speech.

The following objective evaluations were conducted to explore different hyperparameter settings to find the most efficient network architecture. Three design principles were adopted to select the final QPPWG architecture. First, because of the more efficient speech modeling of the QP structure, we try to reduce the number of residual blocks while maintaining a similar speech quality. Secondly, since the {\it receptive field} length is highly related to the speech modeling capacity, if the performance differences are small, the model with the longest {\it receptive field} length will be selected. Last, the motivation of this work is to improve the pitch controllability, so the pitch accuracy is the first priority.

\begin{table}[t]
\caption{CNN Channels of PWG Generator}
\label{tb:cnn_ch}
\fontsize{8pt}{9.6pt}
\selectfont
{%
\begin{tabularx}{\columnwidth}{@{}p{2cm}YYYYY@{}}
\toprule
                     & WD                           & \multicolumn{4}{c}{PWG}   \\ 
Channels             & -                            & 64   & 32   & 16   & 8    \\ \midrule
MCD (dB)             & \cellcolor[HTML]{D9D9D9}2.58 & \textbf{3.69} & 4.15 & 4.23 & 4.89 \\
$F_{0}$RMSE          & \cellcolor[HTML]{D9D9D9}0.10 & \textbf{0.12} & 0.14 & 0.15 & 0.20 \\
$U/V$ (\%)           & \cellcolor[HTML]{D9D9D9}10   & \textbf{14}   & 16   & 16   & 15   \\
Size ($\times 10^6$) & \cellcolor[HTML]{D9D9D9}-    & 1.16 & 0.34 & 0.11 & \textbf{0.04} \\ \bottomrule
\end{tabularx}%
}
\end{table}

\begin{table}[t]
\fontsize{8pt}{9.6pt}
\selectfont
\caption{Blocks and Cycles of PWG Generators with 16 CNN Channels}
\label{tb:pwg_b}
{%
\begin{tabularx}{\columnwidth}{@{}p{2cm}YYYYY@{}}
\toprule
Block Num        & 30   & 20   & 10   & 20   & 20   \\
B$_\mathrm{Fix}$ & $10\times3$ & $10\times2$ & $10\times1$ & $5\times4$ & $20\times1$ \\ \midrule
MCD (dB)                       & \textbf{4.23} & 4.61 & 5.95 & 4.59 & 5.98 \\
$F_{0}$RMSE                    & \textbf{0.15} & 0.17 & 0.31 & 0.35 & 0.30 \\
$U/V$ (\%)                     & \textbf{16}   & 17   & 33   & 44   & 27   \\
Size ($\times 10^6$)           & 0.11 & 0.08 & \textbf{0.04} & 0.08 & 0.08 \\ \bottomrule
\end{tabularx}%
}
\end{table}

\subsubsection{Number of CNN Channels}
To efficiently explore different network architectures and hyperparameter settings, we first explored the relationship between model capacities and the number of CNN channels and tried to reduce the CNN channels for fast model training while keeping reasonable speech quality. The vanilla PWG generators with 8--64 CNN channels were evaluated. Note that because this work focused on improving the generator, all PWG/QPPWG models in this section adopted the same discriminator, whose number of CNN channels was 64 and whose model size was 0.1~M. The results in Table~\ref{tb:cnn_ch} show that the original setting (64 CNN channels) predictably achieves the best performance characteristics of all objective measurements. However, even if the number of CNN channels is reduced to 16, which greatly reduces the training time because of the compact model size, the speech quality and pitch accuracy are still acceptable. Therefore, the objective evaluations in the following sections were conducted based on the models with 16 CNN channels.

\subsubsection{Numbers of Blocks and Cycles}
Since one of the motivations for adopting the QP structure is taking advantage of the higher speech modeling capability to reduce the model size, the importance of the numbers of residual blocks was first evaluated. As shown in Table~\ref{tb:pwg_b}, we first kept 10 residual blocks in one cycle and reduced the number of cycles to cut down the number of total blocks. The results show that the model with 20 blocks still achieves acceptable performance while the performance of the 10 blocks model significantly degrades. Moreover, the importance of the dilation cycle number was also evaluated. The results indicate that compared to the model with two cycles, the four cycles model achieves slightly higher spectral modeling accuracy but much lower pitch accuracy, and both the spectral and pitch accuracies of the one cycle model markedly degrades. In conclusion, although fewer dilation cycles result in a longer {\it receptive field}, the network may not model the speech well. By contrast, the larger the number of dilation cycles, the shorter the {\it receptive field}. Since a longer {\it effective receptive field} can be achieved by replacing fixed blocks with adaptive blocks, we focus on improving the PWG generators with 20 residual blocks and two or four cycles using the QP structure in this paper.

\begin{table}[t]
\fontsize{8pt}{9.6pt}
\selectfont
\caption{Ratios of Fixed and Adaptive Blocks of QPPWG Generators with 20 Residual Blocks, 16 CNN Channels, and Dense Factor 4}
\label{tb:qp_ba}
{%
\begin{tabularx}{\columnwidth}{@{}p{2cm}YYYY@{}}
\toprule
Macro 0 (B$_\mathrm{Ada}$) & $5\times1$ & $5\times2$ & $5\times3$ & $5\times4$ \\ 
Macro 1 (B$_\mathrm{Fix}$) & $5\times3$ & $5\times2$ & $5\times1$ & - \\ \midrule 
                 & \multicolumn{4}{c}{\cellcolor[HTML]{F2F2F2}MCD   (dB)}              \\  
$1\times F_0$    & \textbf{4.79} & \textbf{4.79} & 5.58 & 7.48 \\ 
$1/2\times F_0$  & \textbf{5.22} & 5.29 & 6.03 & 8.16\\ 
$2\times F_0$    & \textbf{5.66} & 6.03 & 7.13 & 8.47 \\ \midrule 
Average          & \textbf{5.22} & 5.37 & 6.24 & 8.04 \\ \midrule 
                 & \multicolumn{4}{c}{\cellcolor[HTML]{F2F2F2}RMSE of log $F_0$}       \\ 
$1\times F_0$    & 0.13 & \textbf{0.12} & 0.13 & 0.14 \\
$1/2\times F_0$  & 0.22 & \textbf{0.17} & \textbf{0.17} & 0.19 \\
$2\times F_0$    & \textbf{0.10} & 0.12 & 0.12 & 0.14 \\ \midrule
Average          & 0.15 & \textbf{0.14} & \textbf{0.14} & 0.15 \\ \midrule
                 & \multicolumn{4}{c}{\cellcolor[HTML]{F2F2F2}$U/V$ decision error (\%)}\\
$1\times F_0$    & 23 & \textbf{16} & \textbf{16} & 20 \\
$1/2\times F_0$  & 26 & 21 & \textbf{20} & 22 \\
$2\times F_0$    & 18 & \textbf{15} & 16 & 18 \\ \midrule
Average          & 23 & \textbf{17} & \textbf{17} & 20 \\ \bottomrule
\end{tabularx}%
}
\end{table}

\subsubsection{Ratio of Fixed and Adaptive Blocks}
Since speech is a quasi-periodic signal, speech modeling is theoretically required both fixed and adaptive blocks to respectively model aperiodic and periodic components. To explore the efficient ratio of fixed and adaptive blocks, four QPPWG models with 20 residual blocks, four cycles, and {\it dense factor} 4 were evaluated. Specifically, because of the more possible combinations of fixed and adaptive blocks, the number of cycles was set to four. The {\it dense factor} was empirically set to 4, and more discussions of {\it dense factor} are presented in the following subsection. As shown in Table~\ref{tb:qp_ba}, although the model with only adaptive blocks (B$_\mathrm{Ada}5\times4$) has the longest {\it receptive fields}, the spectral modeling accuracy is markedly low because of the limited modeling capability of the aperiodic components. The same tendency can also be observed in the spectral domain. The more adaptive blocks the model has, the more harmonic components the generated speech has. However, overenhanced harmonic structures generate significantly robotic and unnatural sounds.

Since the model with balanced numbers of adaptive and fixed blocks achieves the highest pitch accuracy and lowest $U/V$ error while keeping acceptable spectral accuracy and attaining longer {\it receptive fields} than the model with only five adaptive blocks, the 20 residual blocks with balanced numbers of adaptive and fixed blocks was selected as the QPPWG paradigm. To summarize, the ratio of adaptive and fixed blocks is crucial to the network for avoiding over/undermodeling the harmonic structures. Moreover, since one dilation cycle including 10 fixed blocks showed effectiveness in the PWG and WN models, and the {\it receptive fields} of 10 fixed blocks are longer than that of $5\times2$ fixed blocks, the architecture of the following QPPWG models was set to 20 residual blocks including $5\times2$ adaptive blocks and 10 fixed blocks. The QPPWG architecture is denoted as QPPWG\_20.

\begin{table}[t]
\caption{QP Structure of QPPWG\_20 Generator with \protect \\ 16 CNN Channels and Dense Factor 4}
\label{tb:qp_struct}
\fontsize{8pt}{9.6pt}
\selectfont
{%
\begin{tabularx}{\columnwidth}{@{}p{1.5cm}YYYYYY@{}}
\toprule
QP structure & \multicolumn{3}{c}{stacked} & \multicolumn{3}{c}{parallel} \\ \midrule
 & {\cellcolor[HTML]{F2F2F2}MCD}  & {\cellcolor[HTML]{F2F2F2}RMSE} & {\cellcolor[HTML]{F2F2F2}$U/V$} 
 & {\cellcolor[HTML]{F2F2F2}MCD}  & {\cellcolor[HTML]{F2F2F2}RMSE} & {\cellcolor[HTML]{F2F2F2}$U/V$} \\
$1\times F_0$    & \textbf{5.10} & \textbf{0.14} & \textbf{18} & 5.80 & 0.32 & 28 \\
$1/2\times F_0$  & \textbf{5.49} & \textbf{0.18} & \textbf{21} & 6.05 & 0.48 & 43 \\
$2\times F_0$    & 6.32 & \textbf{0.14} & \textbf{26} & \textbf{6.00} & 0.45 & 52 \\ \midrule
Average          & \textbf{5.63} & \textbf{0.15} & \textbf{22} & 5.95 & 0.42 & 41 \\ \bottomrule
\end{tabularx}%
}
\end{table}

\subsubsection{QP Structure}
Since the fixed and adaptive blocks are assumed to respectively model aperiodic and periodic components of speech signals, a new parallel QP structure (Fig.~\ref{fig:qppwg} (b)) was evaluated in this paper compared to the original stacked QP structure (Fig.~\ref{fig:qppwg} (a)). However, the results in Table~\ref{tb:qp_struct} show that the QPPWG\_20 model with a parallel QP structure achieves very low pitch accuracy and high $U/V$ errors, which indicate the very limited periodic component modeling capability of the parallel model.  Observing the output waveforms of the skip connection summation from the adaptive/fixed blocks, we also find that the output waveforms are dominated by the fixed blocks in the parallel QP model while the outputs of the adaptive blocks are very small. In other words, these results show that only the fixed blocks are well activated for speech modeling when the parallel QP structure is adopted. 

The possible reason is that the difficulty of modeling speech using a fixed network architecture is lower than that of the network adopting a more complicated pitch-adaptive architecture in the very initial stage. Since the gradient paths of the fixed and adaptive macroblocks are separated, this difference of modeling difficulty may make the whole adaptive macroblock inactive. On the other hand, because the adaptive and fixed macroblocks are cascaded in the stacked QP structure, these macroblocks are in the same gradient flow, which makes the entire network participates in the speech modeling. Furthermore, since the aperiodic and periodic components are not completely independent, the stacked QP structure takes advantage of the aperiodic and periodic information propagations between the fixed and adaptive macroblocks to get better speech modeling capability. As a result, the stacked QP structure was selected as the QPPWG paradigm. Further discussion and more details about the outputs of the adaptive and fixed macroblocks will be presented in Section~\ref{discussion}. Moreover, the cascaded adaptive to fixed macroblock order is denoted as $af$, and the reversed macroblock order is denoted as $fa$. The effectiveness of the macroblock order will be presented in the overall objective evaluation.

\begin{table}[t]
\caption{Dense Factor of QPPWG${af}$\_20 Generator with \protect \\ 16 CNN Channels}
\label{tb:qp_dense}
\fontsize{8pt}{9.6pt}
\selectfont
{%
\begin{tabularx}{\columnwidth}{@{}p{1.5cm}YYYYY@{}}
\toprule
Dense $a$  & 16           & 8           & 4           & 2           & 1          \\ \midrule
        & \multicolumn{5}{c}{\cellcolor[HTML]{F2F2F2}MCD (dB)}                \\
$1\times F_0$    & 5.26 & 5.26 & \textbf{5.10} & 5.35 & 5.36 \\
$1/2\times F_0$  & 5.64 & 5.57 & \textbf{5.49} & 5.61 & 5.61 \\
$2\times F_0$    & \textbf{5.92} & 6.06 & 6.32 & 5.99 & 6.03 \\ \midrule
Average          & \textbf{5.60} & 5.63 & 5.63 & 5.65 & 5.67\\ \midrule
        & \multicolumn{5}{c}{\cellcolor[HTML]{F2F2F2}RMSE of log $F_0$}          \\
$1\times F_0$    & \textbf{0.13} & \textbf{0.13} & 0.14 & 0.14 & 0.17 \\
$1/2\times F_0$  & 0.21 & \textbf{0.17} & 0.18 & 0.23 & 0.28 \\
$2\times F_0$    & 0.14 & 0.14 & 0.14 & 0.14 & 0.15 \\ \midrule
Average          & 0.16 & \textbf{0.14} & 0.15 & 0.17 & 0.20\\ \midrule
        & \multicolumn{5}{c}{\cellcolor[HTML]{F2F2F2}$U/V$ decision error (\%)} \\
$1\times F_0$    & 17 & 17 & 18 & 17 & 17 \\
$1/2\times F_0$  & 25 & \textbf{21} & \textbf{21} & 24 & 27 \\
$2\times F_0$    & 24 & \textbf{19} & 26 & 20 & 20 \\ \midrule
Average          & 22 & \textbf{19} & 22 & 20 & 21 \\ \bottomrule
\end{tabularx}%
}
\end{table}

\subsubsection{Dense Factor}
The {\it dense factor} is inversely proportional to the {\it receptive field} size, and the QPPWG${af}$\_20 models with 1--16 {\it dense factors} were evaluated. The results in Table~\ref{tb:qp_dense} show that while the models with {\it dense factors} of 4--16  achieve similar generative performance, the models with {\it dense factors} of 1 and 2 achieve slightly worse performance. A similar tendency was also observed by listening to the generated speech. The generated utterances from the models with {\it dense factors} of 1 and 2 were more unstable. Furthermore, PDCNN degenerates to DCNN when $E_{t}$ is one, and a larger {\it dense factor} makes $E_{t}$ closer to one for more $F_{0}$ values. Therefore, since a lower {\it dense factor} attains a longer {\it receptive field} expansion and a higher lower bound of $F_{0}$, which makes PDCNN degenerate to DCNN, the {\it dense factors} of the following QPPWG models were set to 4.  

\begin{table}[t]
\caption{Model Size ($G$: Generator; $D$: Discriminator)}
\label{tb:model_size}
{%
\begin{tabularx}{\columnwidth}{@{}p{2cm}YYYYY@{}}
\toprule
                           & QPNet & \multicolumn{3}{c}{PWG} \\ 
Block Num                  & 16 & 30 & 20 & 16 \\
Macro 0 (B$_\mathrm{Fix}$) & $4\times3$ & $10\times3$ & $10\times2$ & $4\times4$ \\
Macro 1 (B$_\mathrm{Ada}$) & $4\times1$  & -      & -      & -      \\ \midrule
$G$ ($\times 10^6$) & 24 & 1.16 & 0.78 & 0.63 \\
$D$ ($\times 10^6$) & -  & 0.10 & 0.10 & 0.10 \\ \midrule
         & \multicolumn{2}{c}{QPPWG$af$} & \multicolumn{2}{c}{QPPWG$fa$} \\
Macro order 
& \multicolumn{2}{c}{B$_\mathrm{Ada}\rightarrow$ B$_\mathrm{Fix}$} 
& \multicolumn{2}{c}{B$_\mathrm{Fix}\rightarrow$ B$_\mathrm{Ada}$} \\
Block Num & 20 & 16 & 20 & 16 \\
Macro 0  & $5\times2$  & $4\times2$ & $10\times1$  & $4\times2$ \\
Macro 1  & $10\times1$ & $4\times2$ & $5\times2$ & $4\times2$ \\  \midrule
$G$ ($\times 10^6$) & 0.79 & 0.63 & 0.79 & 0.63 \\
$D$ ($\times 10^6$) & 0.10 & 0.10 & 0.10 & 0.10 \\ \bottomrule
\end{tabularx}%
}
\end{table}

\begin{table*}[t]
\caption{WORLD, QPNet, PWG, and QPPWG Vocoders}
\label{tb:overall}
\fontsize{8pt}{9.6pt}
\selectfont
{%
\begin{tabularx}{2.0\columnwidth}{@{}p{1.3cm}YYYYYYYYYYYYYYYYYY@{}}
\toprule
Gender & \multicolumn{9}{c}{\cellcolor[HTML]{F2F2F2}Male} & \multicolumn{9}{c}{Female} \\ \midrule
Vocoder 
& WD & QPNet & \multicolumn{3}{c}{PWG} & \multicolumn{2}{c}{QPPWG$af$} & \multicolumn{2}{c}{QPPWG$fa$} 
& WD & QPNet & \multicolumn{3}{c}{PWG} & \multicolumn{2}{c}{QPPWG$af$} & \multicolumn{2}{c}{QPPWG$fa$} \\
Block Num & - & 16 & 30 & 20 & 16 & 20 & 16 & 20 & 16 & - & 16 & 30 & 20 & 16 & 20 & 16 & 20 & 16 \\ \midrule
       & \multicolumn{18}{c}{\cellcolor[HTML]{F2F2F2}MCD (dB)} \\
$1\times F_0$ 
& \cellcolor[HTML]{D9D9D9}{2.57} & 4.29 & \textbf{3.61} & 3.70 & 4.28 & 3.72 & 4.15 & 4.44 & 5.01 
& \cellcolor[HTML]{D9D9D9}{2.59} & 4.11 & \textbf{3.76} & 3.79 & 4.21 & 3.87 & 4.20 & 4.65 & 4.98 \\
$1/2\times F_0$ 
& \cellcolor[HTML]{D9D9D9}{5.09} & 5.29 & 5.09 & \textbf{4.90} & 5.02 & 5.08 & 5.41 & 5.49 & 5.82 
& \cellcolor[HTML]{D9D9D9}{2.69} & 4.55 & \textbf{3.85} & 3.88 & 4.28 & 3.95 & 4.36 & 4.88 & 5.37 \\
$2\times F_0$ 
& \cellcolor[HTML]{D9D9D9}{3.10} & 4.23 & 4.13 & \textbf{4.10} & 4.39 & 4.12 & 4.58 & 4.78 & 5.21 
& \cellcolor[HTML]{D9D9D9}{4.49} & 4.99 & 6.34 & 6.03 & \textbf{4.72} & 5.73 & 6.26 & 6.44 & 6.73 \\
\midrule
Average
& \cellcolor[HTML]{D9D9D9}{3.59} & 4.60 & 4.28 & \textbf{4.23} & 4.56 & 4.31 & 4.71 & 4.90 & 5.35 
& \cellcolor[HTML]{D9D9D9}{3.26} & 4.55 & 4.65 & 4.57 & \textbf{4.41} & 4.52 & 4.94 & 5.32 & 5.69 \\
\midrule
& \multicolumn{18}{c}{\cellcolor[HTML]{F2F2F2}RMSE of log $F_0$} \\
$1\times F_0$    
& \cellcolor[HTML]{D9D9D9}{0.13} & 0.19 & 0.15 & 0.19 & 0.54 & 0.15 & \textbf{0.13} & 0.14 & 0.16 
& \cellcolor[HTML]{D9D9D9}{0.07} & 0.09 & 0.08 & 0.10 & 0.28 & \textbf{0.07} & 0.08 & \textbf{0.07} & 0.08 \\
$1/2\times F_0$    
& \cellcolor[HTML]{D9D9D9}{0.20} & 0.35 & 0.43 & 0.53 & 0.54 & 0.30 & \textbf{0.21} & 0.30 & 0.29 
& \cellcolor[HTML]{D9D9D9}{0.08} & 0.11 & 0.11 & 0.11 & 0.29 & \textbf{0.09} & \textbf{0.09} & \textbf{0.09} & \textbf{0.09} \\
$2\times F_0$    
& \cellcolor[HTML]{D9D9D9}{0.12} & 0.16 & 0.14 & 0.14 & 0.39 & 0.13 & \textbf{0.11} & 0.12 & 0.12 
& \cellcolor[HTML]{D9D9D9}{0.09} & 0.20 & 0.15 & 0.16 & 1.08 & 0.10 & \textbf{0.09} & 0.11 & 0.10 \\ \midrule
Average    
& \cellcolor[HTML]{D9D9D9}{0.15} & 0.24 & 0.24 & 0.29 & 0.49 & 0.19 & \textbf{0.15} & 0.19 & 0.19 
& \cellcolor[HTML]{D9D9D9}{0.08} & 0.13 & 0.12 & 0.12 & 0.55 & 0.09 & \textbf{0.08} & 0.09 & 0.09 \\
\midrule
& \multicolumn{18}{c}{\cellcolor[HTML]{F2F2F2}$U/V$ decision error (\%)} \\
$1\times F_0$    
& \cellcolor[HTML]{D9D9D9}{12} & \textbf{17} & 20 & 19 & 53 & 20 & 21 & 19 & 22 
& \cellcolor[HTML]{D9D9D9}{8} & 11 & \textbf{9} & 10 & 57 & 13 & 14 & 11 & 10 \\
$1/2\times F_0$    
& \cellcolor[HTML]{D9D9D9}{17} & 32 & 23 & 23 & 38 & 23 & \textbf{22} & 23 & \textbf{22} 
& \cellcolor[HTML]{D9D9D9}{12} & 20 & \textbf{19} & 20 & 52 & 22 & 23 & 22 & 23 \\
$2\times F_0$    
& \cellcolor[HTML]{D9D9D9}{11} & 17 & \textbf{12} & 13 & 62 & 16 & 18 & 14 & 13 
& \cellcolor[HTML]{D9D9D9}{11} & 28 & 13 & 21 & 69 & 23 & 11 & 12 & \textbf{10} \\ 
\midrule
Average    
& \cellcolor[HTML]{D9D9D9}{13} & 22 & \textbf{18} & \textbf{18} & 51 & 20 & 20 & 19 & 19 
& \cellcolor[HTML]{D9D9D9}{10} & 20 & \textbf{14} & 17 & 60 & 19 & 16 & 15 & \textbf{14} \\ 
\bottomrule
\end{tabularx}%
}
\end{table*}

\subsubsection{Overall Objective Evaluation}
An overall objective evaluation was conducted including the WD, QPNet, PWG, and QPPWG models. Specifically, since the AR QP structure has shown effectiveness for the WN~\cite{qpnet_2019, qpnet_2020} vocoder, it is interesting to explore the generality of the QP structure for non-AR models and the performance difference between the QPNet and QPPWG models. Because the QPNet architecture contained only 16 residual blocks with four cycles, the PWG and QPPWG models with 16 residual blocks and four cycles were also evaluated. Moreover, the effectiveness of the different QPPWG macroblock orders was also explored. The number of CNN channels of the PWG and QPPWG models was set to 64 following the original setting. The model sizes are shown in Table~\ref{tb:model_size}. Since the model size is proportional to the square of the number of CNN channels, the model size of vanilla PWG is only 5~\% of that of QPNet because of the greatly reduced number of CNN channels. The sizes of the QPPWG models were reduced further by 30--50~\% because of the reduced number of residual blocks compared with that of vanilla PWG (PWG\_30).

To present the correlations of the $F_0$ distributions and vocoder performances, the gender-dependent results are shown in Table~\ref{tb:overall}. Specifically, because of the multi-speaker training manner, the $F_0$ range of the training data covered both male and female $F_0$ values. Therefore, the most female $1/2\times F_0$ and male $2\times F_0$ values are still in the $F_0$ range of the training data while the most female $2\times F_0$ and male $1/2\times F_0$ values are outside the $F_0$ range. Since these gender-dependent differences might cause different effects in the scaled $F_0$ evaluations, the gender-dependent results are more informative.

As the MCD results shown in Table~\ref{tb:overall}, the female $2\times F_0$ and male $1/2\times F_0$ sets achieve much higher MCD than the female $1/2\times F_0$ and male $2\times F_0$ sets as we expected. However, the overall tendencies of the male and female sets are similar. The QPPWG models with the ${af}$ order outperform the models with ${fa}$ order in both sets and all scenarios showing the superiority of QPPWG$af$. The possible reason is that modeling the long-term structure of speech signals first as QPPWG$af$ makes the generated speech more stable than modeling the details first as QPPWG$fa$. More details about the comparison between QPPWG$af$ and QPPWG$fa$ will be presented in Section~\ref{discussion}.

Furthermore, the QPPWG${af}$\_20 model achieves a comparable spectral accuracy with the PWG\_30 and PWG\_20 models showing the QP structure keeping the similar spectral prediction accuracy. 
Although the average MCD of PWG\_16-generated utterances is not very high, the very high RMSE of log $F_{0}$ and the very high $U/V$ error indicate that the speech quality of PWG\_16 is low. Specifically, the similar MCDs of PWG\_16-generated utterances with different scaled $F_{0}$ values imply that the PWG\_16 model tends to ignore the $F_{0}$ scaled ratio to generate similar speech waveforms. The very high RMSE of log $F_{0}$ and the very high $U/V$ error also indicate that the PWG\_16-generated speech waveforms lack fine harmonic structures. On the other hand, compared to QPNet, although the model size of QPPWG$af$\_16 is much smaller than that of QPNet, the non-AR mechanism and GAN structure still make QPPWG achieve comparable spectral prediction accuracy.

Because the GAN structure greatly improves the speech modeling capability, the results of the $F_{0}$ RMSE and $U/V$ error in Table~\ref{tb:overall} show that the non-AR PWG models already achieve a comparable pitch accuracy with the AR QPNet model. However, the QP structure still further improves the pitch accuracy of the non-AR PWG models. The QPPWG${af}$\_16 model even attains a similar pitch accuracy to the reference WD vocoder. Although the pitch and $U/V$ accuracies of PWG\_16 markedly degrade because of the short {\it receptive field}, the QPPWG${af}$\_16 model significantly improves them to an acceptable level showing the effectiveness of the QP structure to enlarge {\it receptive filed}. In conclusion, the QP structure efficiently increases the {\it effective receptive field} size and introduces the pitch information to the network, resulting in a comparable spectral accuracy, a much higher pitch accuracy, and a smaller model size. The objective results show the effectiveness of the proposed QP structure for the PWG models. 

\begin{figure*}[ht]
\centering
\centerline{\includegraphics[width=2\columnwidth]{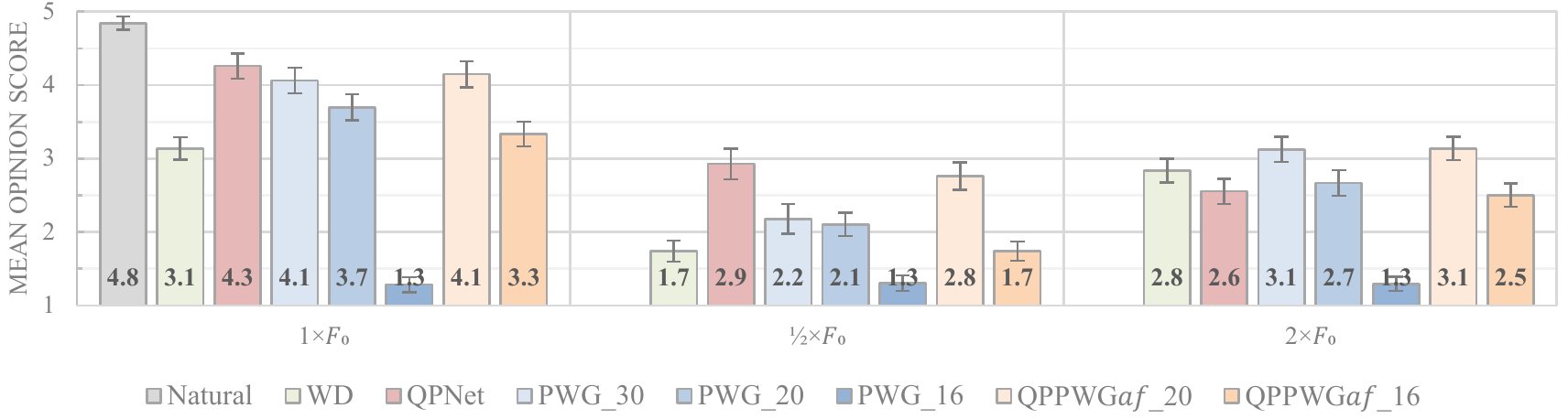}}
\caption{Speech quality MOS evaluations of male speakers with 95~\% CI.}
\label{fig:mos_male}

{\ }

\centering
\centerline{\includegraphics[width=2\columnwidth]{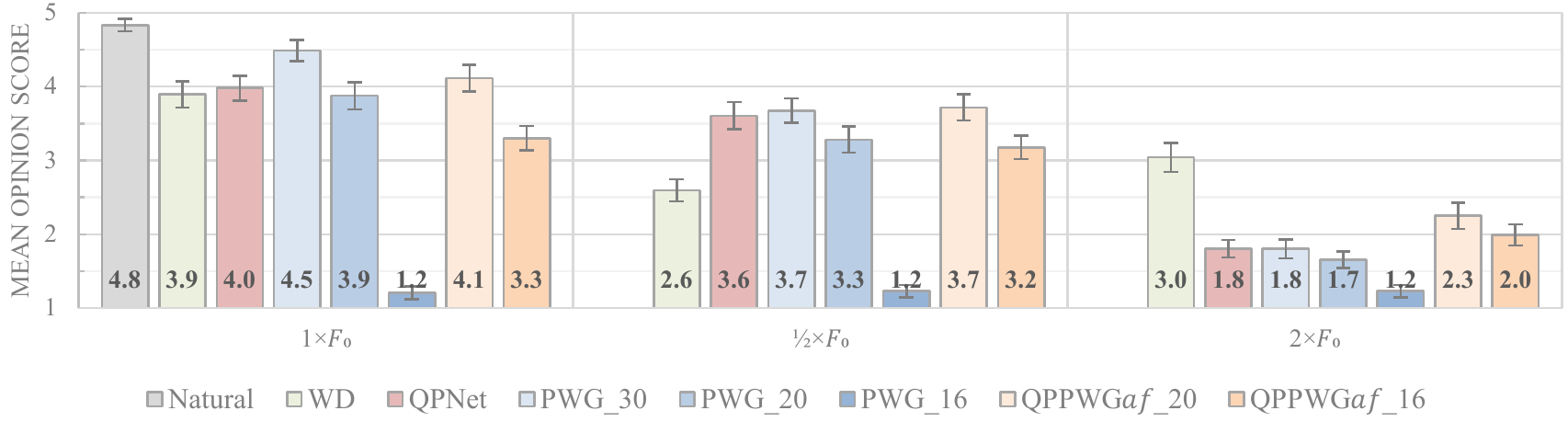}}
\caption{Speech quality MOS evaluations of female speakers with 95~\% CI.}
\label{fig:mos_female}
\end{figure*}

On the other hand, since the WD-extracted $mcep$ and $F_{0}$ are not completely independent, taking $mcep$ extracted from natural speech as the ground truth of the scaled $F_0$ scenarios might cause some mismatches. However, the objective evaluations still provide meaningful information about the performance of these vocoders, and we also conducted the subjective evaluation in the following subsection to provide convincing results from different aspects.

\subsection{Subjective Evaluations}

The set of samples used for subjective evaluation was composed of 1680 synthesized and 80 natural utterances. The synthesized utterances were generated by seven vocoders conditioned on three $F_{0}$ scaled ratios (unchanged, halved, and doubled) and four speakers (the VCC2018 SPOKE set). For each vocoder, speaker, and $F_{0}$ scaled ratio, we randomly selected 20 utterances from the 35 testing utterances for both mean opinion score (MOS) and ABX evaluations. Specifically, the speech quality of each utterance was evaluated by listeners assigning MOSs of 1--5. The higher the MOS, the better the speech quality. For each ABX, two testing utterances were compared with one reference, and the listeners chose the one whose pitch was more consistent with that of the reference. Eight listeners evaluated part of the subjective evaluation set in both MOS and ABX tests, and each utterance/pair was evaluated by at least two listeners. Although the listeners were not native English speakers, they worked on audio-related research. The demo utterances can be found on the demo page~\cite{demo}.

\subsubsection{MOS Evaluation of Speech Quality}
The MOS evaluation included the vocoders of WD, QPNet, PWG of three different sizes, and QPPWG of two different sizes. The MOS results shown in Figs.~\ref{fig:mos_male} and~\ref{fig:mos_female} are presented for three different $F_{0}$ scaled ratios for male and female speakers, respectively. The overall results show that the proposed QP structure improves the speech modeling capacity of the PWG vocoders, especially when the PWG\_16 vocoder has a very small {\it receptive field}. Because the QPPWG vocoders markedly outperform the PWG vocoders of the same size for all scenarios in the MOS evaluation, the following discussion focuses on comparisons among QPPWG${af}$\_20, PWG\_30, and QPNet.

For the $1/2\times F_{0}$ scenario, the QPPWG${af}$\_20 vocoder markedly outperforms the PWG\_30 and WD vocoders and attains a similar speech quality to the QPNet vocoder for the male set. For the female set, the QPPWG${af}$\_20 vocoder is comparable to the PWG\_30 and QPNet vocoders while still outperforming the WD vocoder. The results indicate that the models with the QP structure are more robust for an unseen $F_{0}$ outside the $F_{0}$ range of the training data, such as most of the $1/2\times F_{0}$ values in the male set. On the other hand, although the combination of the $1/2\times F_{0}$ and other acoustic features in the female set is still unseen, the scaled $F_{0}$ values are almost in the $F_{0}$ range of the training data. Therefore, the PWG\_30 vocoder can still achieve a similar speech quality to the QPPWG${af}$\_20 vocoder.

For the $2\times F_{0}$ scenario, because most of the scaled $F_{0}$ values of the male set are in the $F_{0}$ range of the training data, the performance of the QPPWG${af}$\_20 vocoder is similar to that of the PWG\_30 vocoder for the male set. The QPPWG${af}$\_20 vocoder outperforms the WD and QPNet vocoders in the male set, while the QPNet vocoder achieves an inferior speech modeling capacity for the $2\times F_{0}$ scenario~\cite{qpnet_2019, qpnet_2020}. On the other hand, although the QPPWG${af}$\_20 vocoder predictably outperforms the PWG\_30 and QPNet vocoders in the female $2\times F_{0}$ scenario, the WD vocoder achieves a higher speech quality than the QPPWG${af}$\_20 vocoder. A possible reason for this is that many PDCNNs of the QPPWG${af}$\_20 model might degenerate to DCNNs because of the values of $E_{t}$ close to one due to the very high $F_{0}$ values.

In conclusion, the proposed QPPWG vocoder with 20 residual blocks attains speech quality competitive with the PWG vocoder with 30 residual blocks for natural auxiliary features even though the model size is only 70~\% of that of the PWG model. When conditioned on the auxiliary features with the unseen $F_{0}$ values, which are outside the $F_{0}$ range of the training data, the proposed QPPWG vocoders achieve a higher speech quality than the PWG vocoders. The results confirm the effectiveness of the proposed QP structure for the PWG model in efficiently modeling speech signals and dealing with unseen $F_{0}$ features. 

\begin{figure}
\centering
\centerline{\includegraphics[width=0.9\columnwidth]{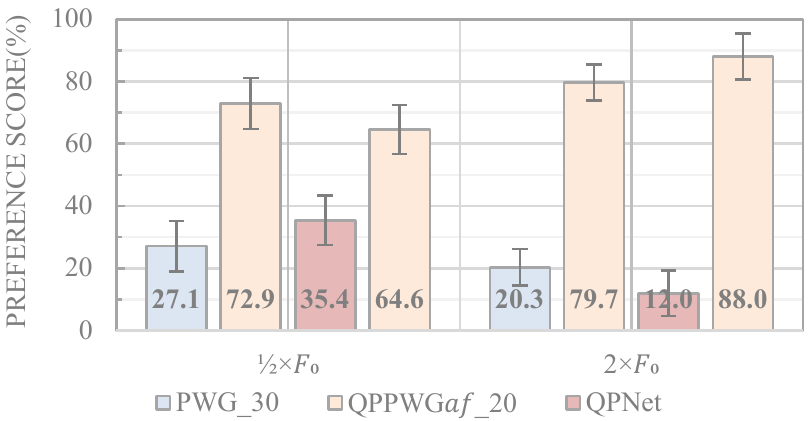}}
\caption{Pitch accuracy ABX evaluations with 95~\% CI.}
\label{fig:abx}
\end{figure}

\subsubsection{ABX Evaluation of Pitch Accuracy}
To evaluate the perceptual pitch accuracy, we conducted ABX tests of the QPPWG${af}$\_20, PWG\_30, and QPNet vocoders with the WD-generated utterances taken as references. Note that because there were no natural utterances with scaled $F_{0}$ and the conventional signal-processing-based vocoder usually attains accurate pitch controllability, the WD-generated utterances were an alternative ground truth. Since the speech quality of the WD-generated speech is usually worse than the neural-vocoder-generated-speech, we asked the listeners to focus on the pitch differences and ignore the speech quality differences. Because the results of the female and male sets have the same tendency, only the overall results are shown in Fig.~\ref{fig:abx}. We find that the perceptual pitch accuracy of the proposed QPPWG${af}$\_20 vocoder is much better than that of the PWG and QPNet vocoders for both halved and doubled $F_{0}$ scenarios. To summarize, the ABX results show perceptible pitch differences between QPPWG- and PWG-/QPNet-generated utterances, and the ABX experimental results are consistent with the objective results of the RMSE of log $F_{0}$

\section{Discussion} \label{discussion}

\subsection{Effective Receptive Field}
\begin{figure}[t]
\centering
\centerline{\includegraphics[width=0.9\columnwidth]{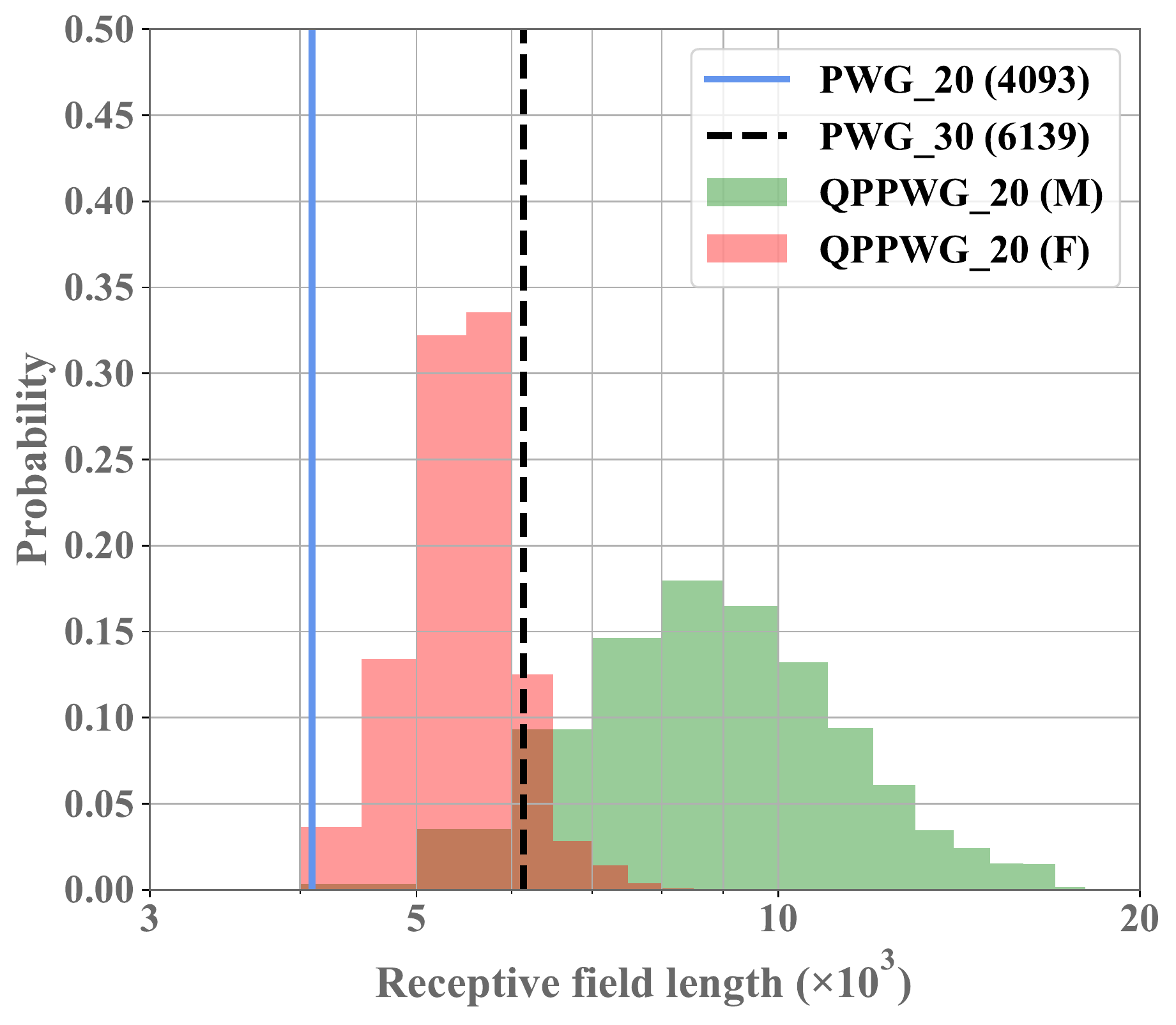}}
\caption{Comparison of receptive field lengths of PWG\_30, PWG\_20, and QPPWG\_20 for male (M) and female (F) sets.}
\label{fig:rp_field}
\end{figure}

Our previous works~\cite{qpnet_2019, qpnet_2020} showed that the capacity of an AR vocoder is strongly related to the length of its {\it receptive field}, and we argue that a non-AR vocoder has a similar tendency. Specifically, the {\it receptive field} length of PWG\_30 is 6139 ($2^{0}+\cdots +2^{9}=1023$ with three cycles and two sides plus one) and that of PWG\_20 is 4093. For the QPPWG, the {\it effective receptive field} length is the summation of 2047 for B$_\mathrm{Fix}10\times1$ and 124$\times E_{t}$ ($2^{0}+\cdots +2^{4}=31$ with two cycles and two sides) for B$_\mathrm{Ada}5\times2$. The male $F_{0}$ range is around 40--240 Hz and the female $F_{0}$ range is around  100--400 Hz, so the $E_{t}$ of the male set is around 20--140 and that of the female set is around 10-–60 when the {\it dense factor} is set to 4. As shown in Fig.~\ref{fig:rp_field}, most of the {\it effective receptive filed} lengths of QPPWG\_20 for the male set are longer than the {\it receptive filed} length of PWG\_30, which may result in the higher pitch accuracy and comparable speech quality of QPPWG. The slightly lower speech quality of QPPWG\_20 than of PWG\_30 for the female set may result from the shorter {\it effective receptive fields} of QPPWG\_20. In conclusion, the quality of the non-AR-vocoder-generated speech still strongly depends on the length of the {\it receptive field}, and QPPWG has longer {\it effective receptive fields} by skipping some redundant samples of the periodic components. Although the network may also lose some details of the aperiodic components owing to the skipping mechanism, the overall experimental results still show the effectiveness of the QP structure.

\begin{table}[t]
\caption{Comparison of RTF of Model Inference}
\label{tb:rtf}
\fontsize{8pt}{9.6pt}
\selectfont
{%
\begin{tabularx}{\columnwidth}{@{}p{2.8cm}YYY@{}}
\toprule
                     & PWG\_20 & PWG\_30 & QPPWG\_20 \\ \midrule
Intel Xeon Gold 6142 & \textbf{0.474}   & 0.579   & 0.512     \\
Nvidia TITAN V       & \textbf{0.011}   & 0.016   & 0.020     \\ \bottomrule
\end{tabularx}%
}
\end{table}

\begin{figure*}[t]
\fontsize{8pt}{9pt}
\selectfont
{%
\begin{tabularx}{2.0\columnwidth}{@{}p{2cm}XXXX@{}}
\toprule
&  \centering\arraybackslash{1--5 blocks}
&  \centering\arraybackslash{1--10 blocks}
&  \centering\arraybackslash{1--15 blocks}
&  \centering\arraybackslash{1--20 blocks} \\ \midrule
  PWG\_20
& (a) $5\times$B$_\mathrm{Fix}$ & (b) $10\times$B$_\mathrm{Fix}$ 
& (c) $15\times$B$_\mathrm{Fix}$ & (d) $20\times$B$_\mathrm{Fix}$\\ 
&  \includegraphics[width=0.39\columnwidth]{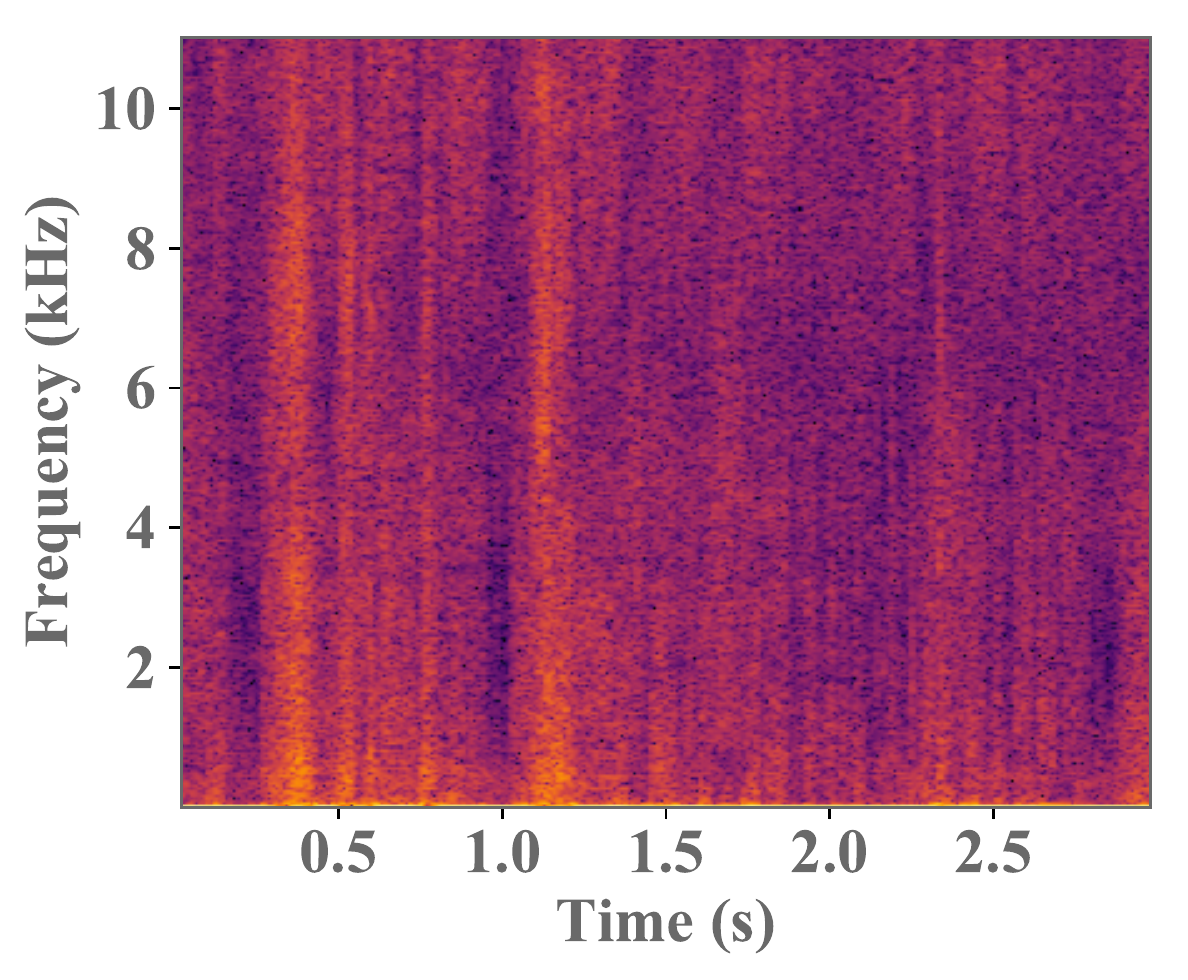}
&  \includegraphics[width=0.39\columnwidth]{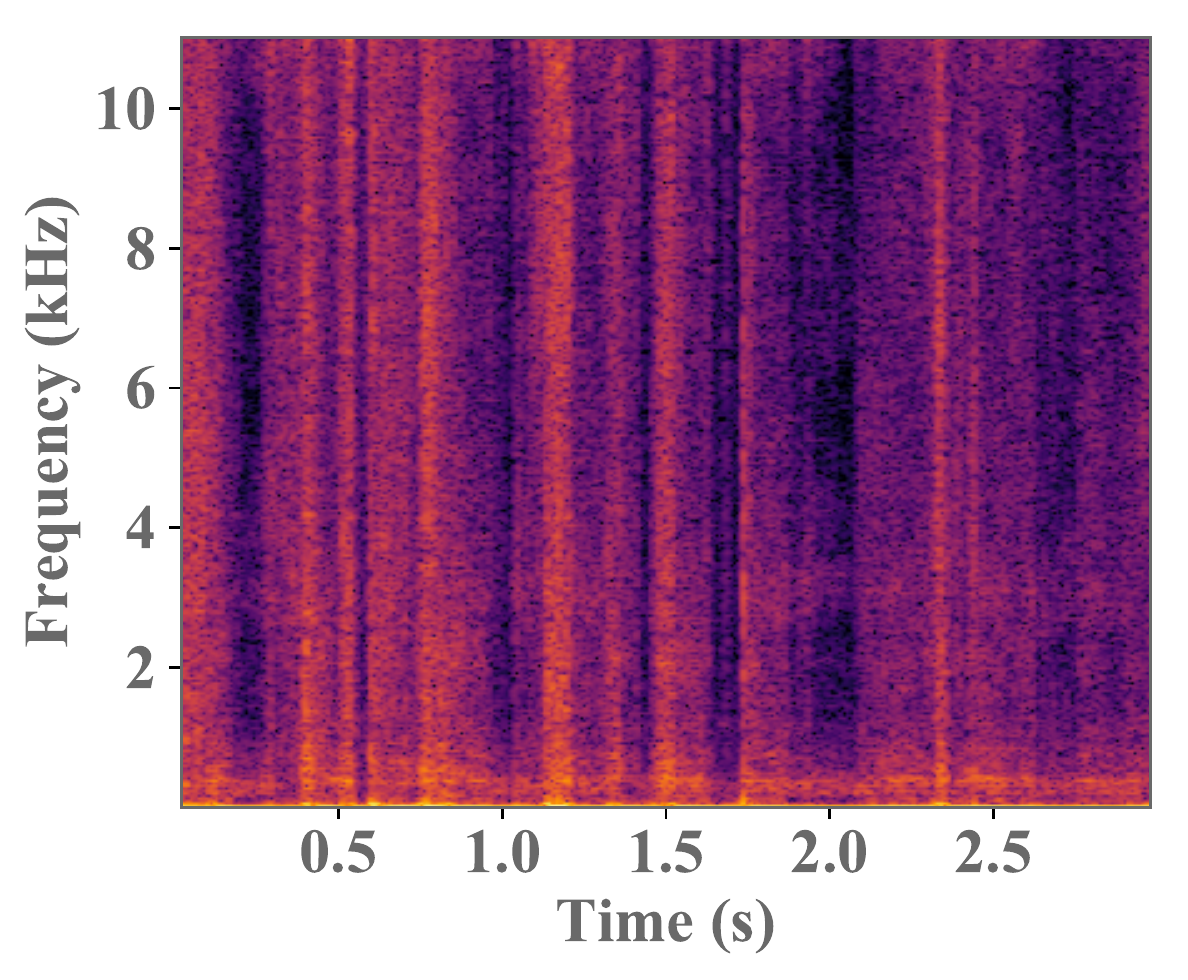}
&  \includegraphics[width=0.39\columnwidth]{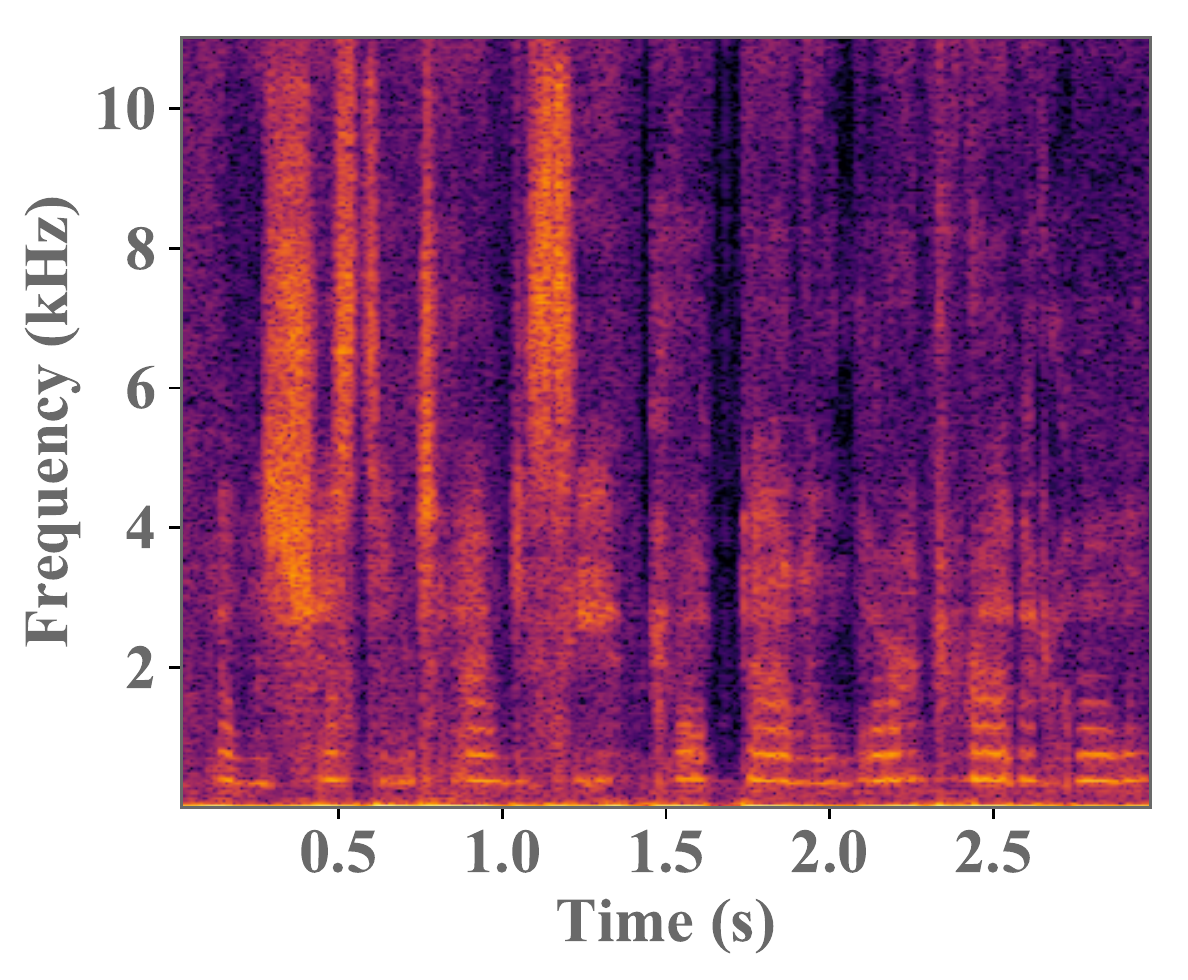}
&  \includegraphics[width=0.39\columnwidth]{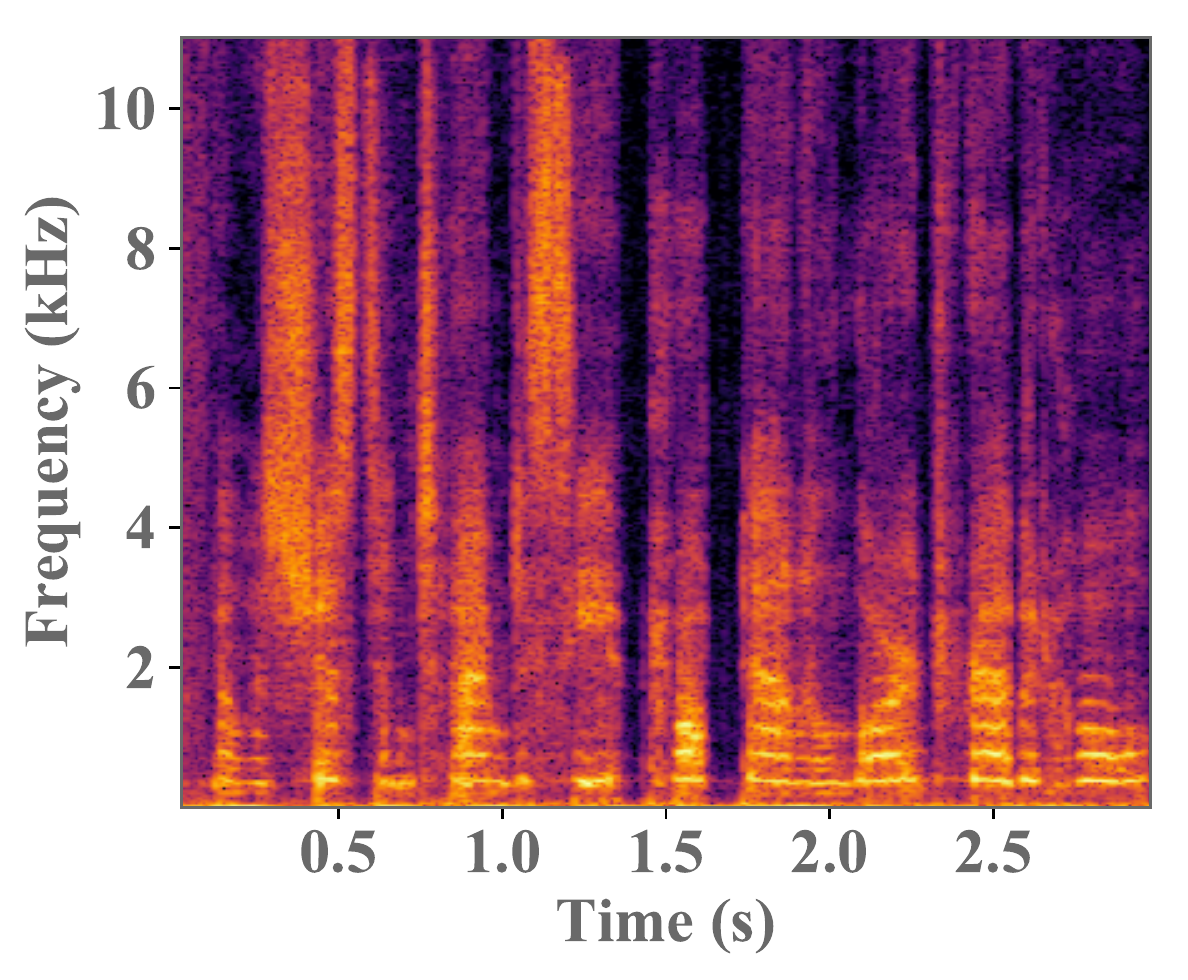} \\ \midrule
 QPPWG$af$\_20 
& (e) $5\times$B$_\mathrm{Ada}$ 
& (f) $10\times$B$_\mathrm{Ada}$ 
& (g) $10\times$B$_\mathrm{Ada}+5\times$B$_\mathrm{Fix}$ 
& (h) $10\times$B$_\mathrm{Ada}+10\times$B$_\mathrm{Fix}$ \\ 

& \includegraphics[width=0.39\columnwidth]{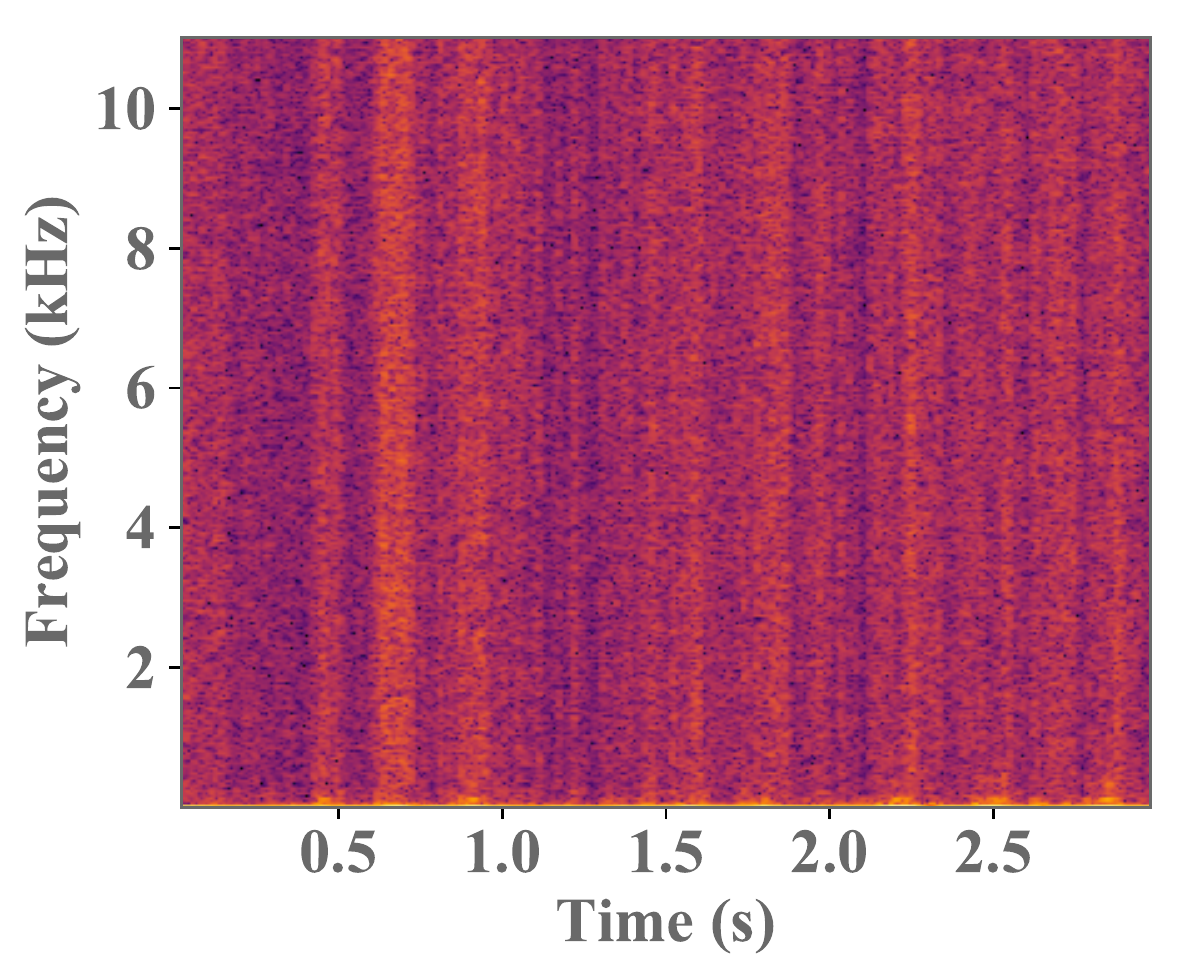}
& \includegraphics[width=0.39\columnwidth]{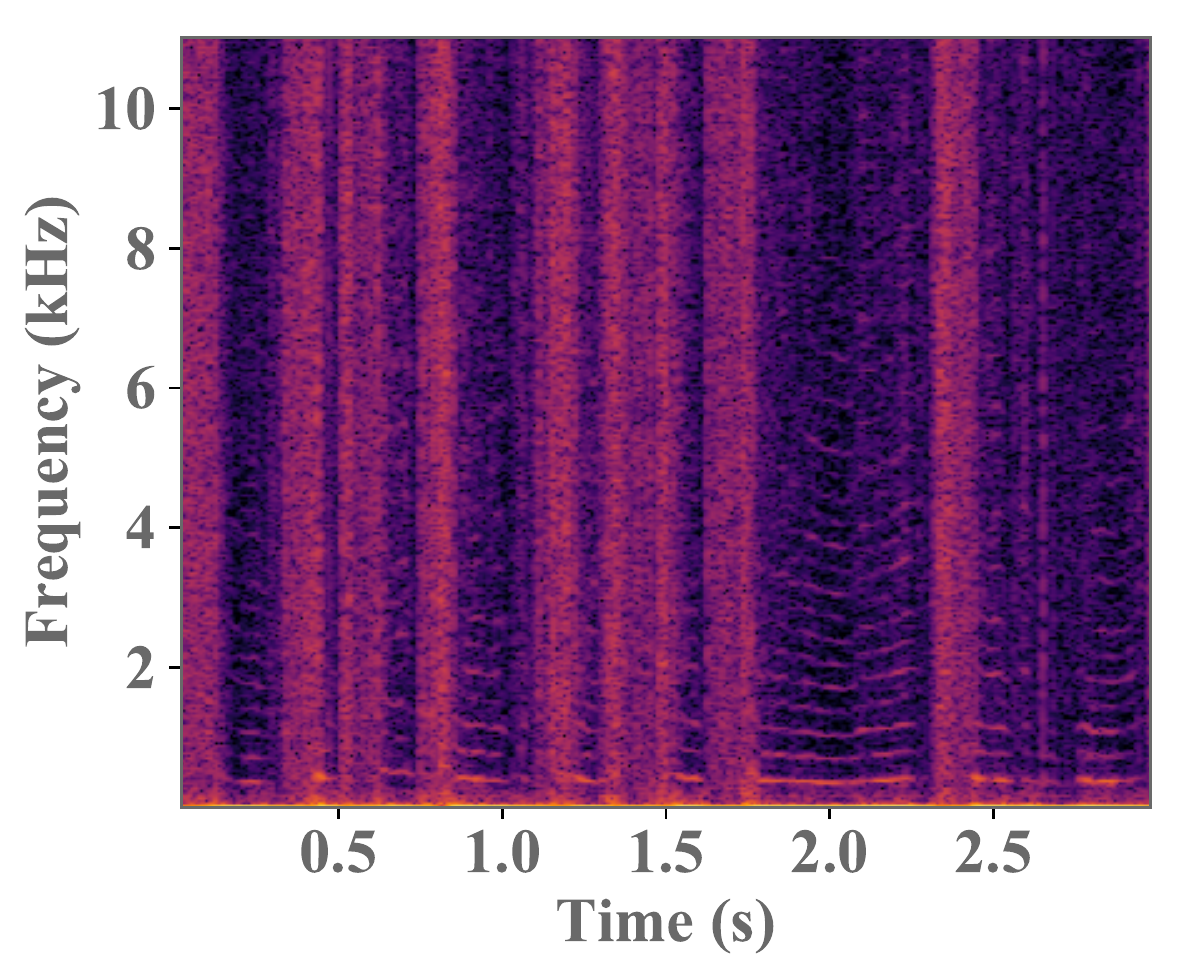} 
& \includegraphics[width=0.39\columnwidth]{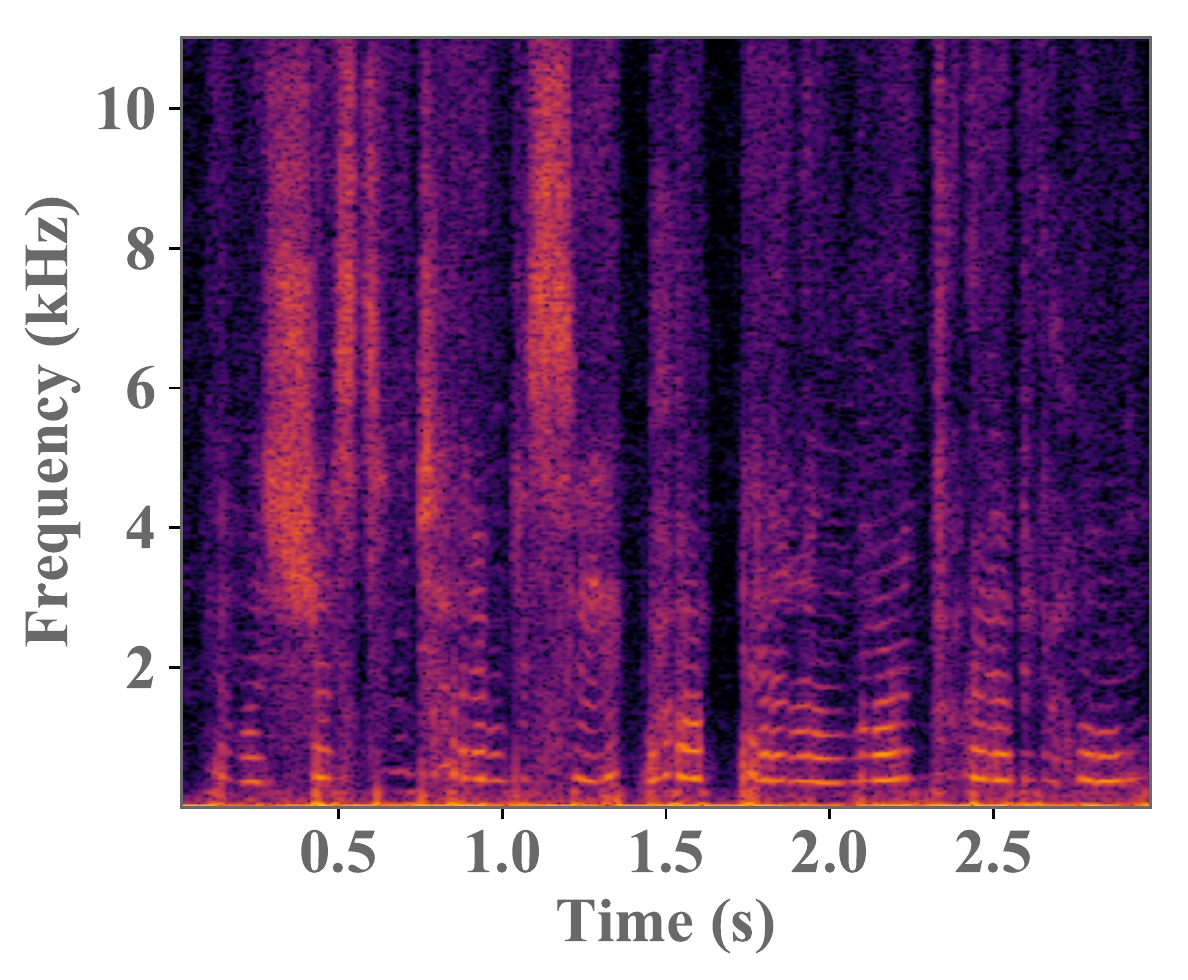} 
& \includegraphics[width=0.39\columnwidth]{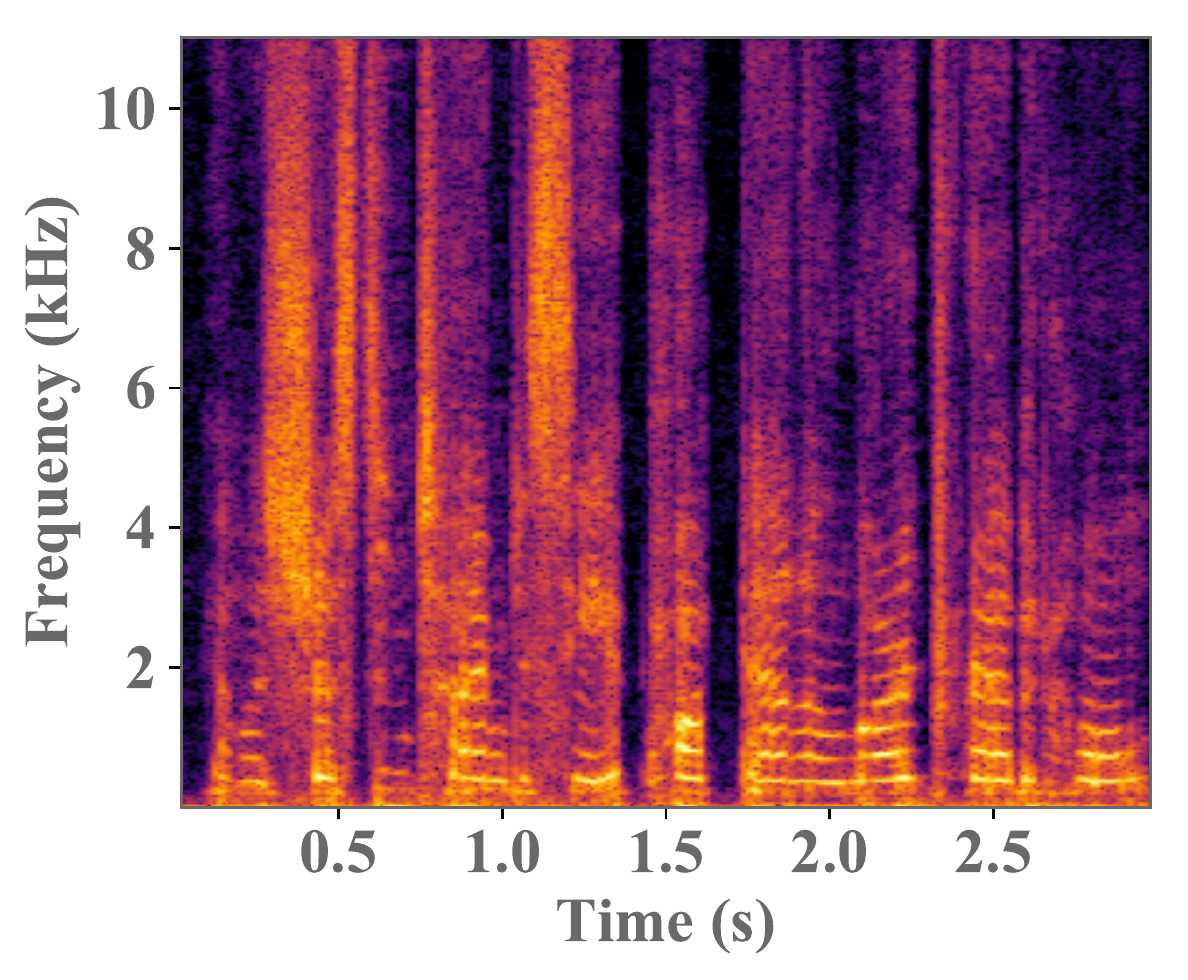} \\ \midrule
  QPPWG$fa$\_20
& (i) $5\times$B$_\mathrm{Fix}$ 
& (j) $10\times$B$_\mathrm{Fix}$ 
& (k) $10\times$B$_\mathrm{Fix}+5\times$B$_\mathrm{Ada}$ 
& (l) $10\times$B$_\mathrm{Fix}+10\times$B$_\mathrm{Ada}$ \\ 
& \includegraphics[width=0.39\columnwidth]{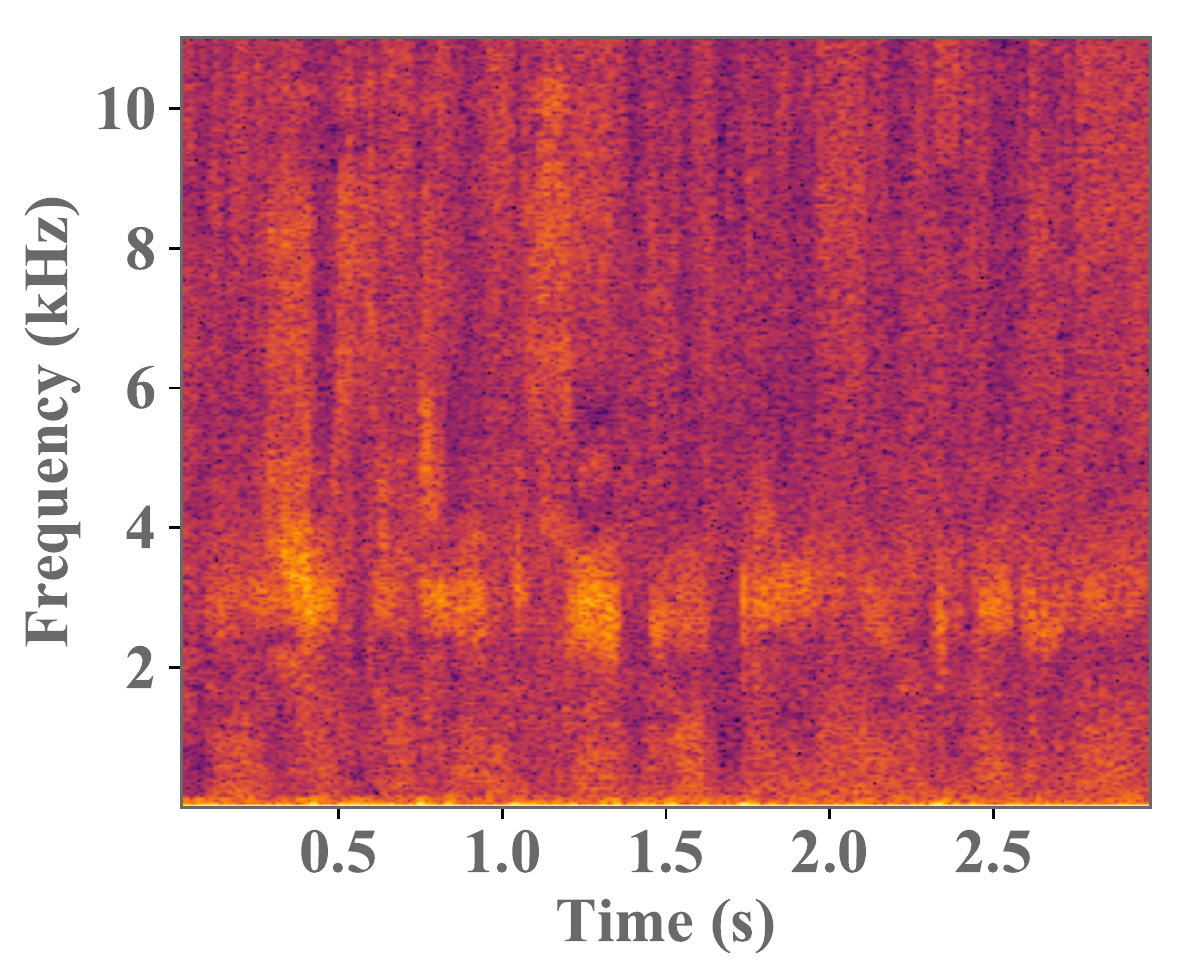}
& \includegraphics[width=0.39\columnwidth]{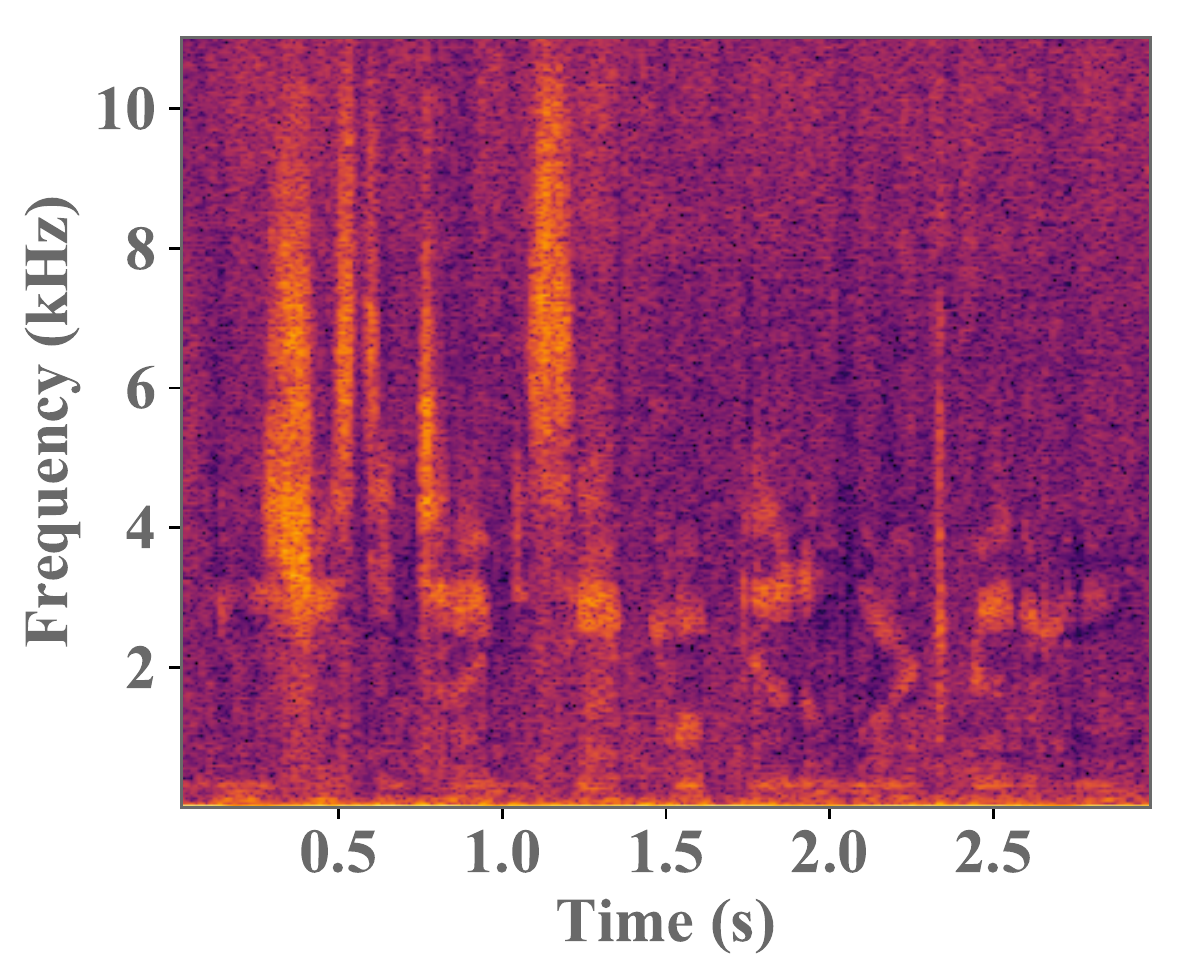} 
& \includegraphics[width=0.39\columnwidth]{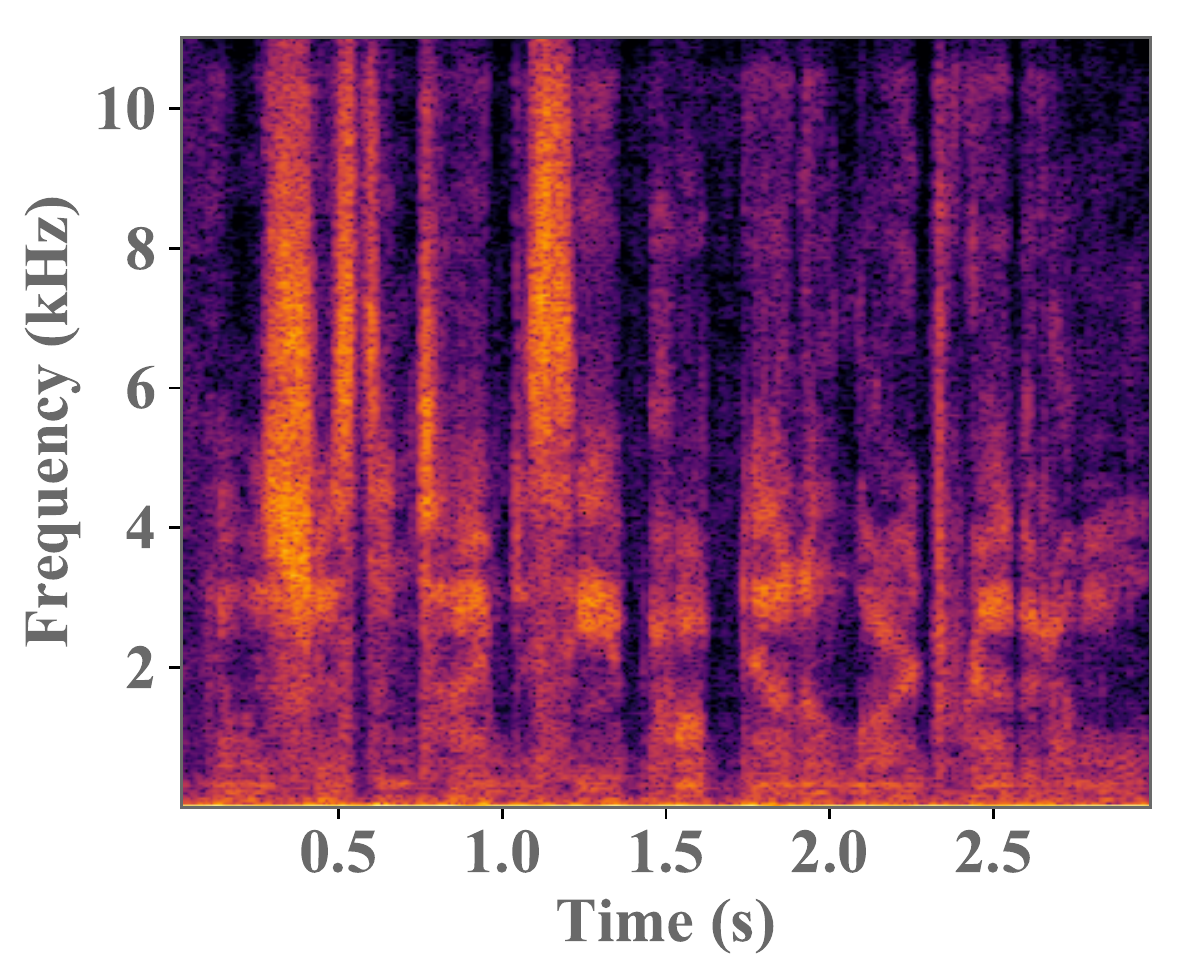} 
& \includegraphics[width=0.39\columnwidth]{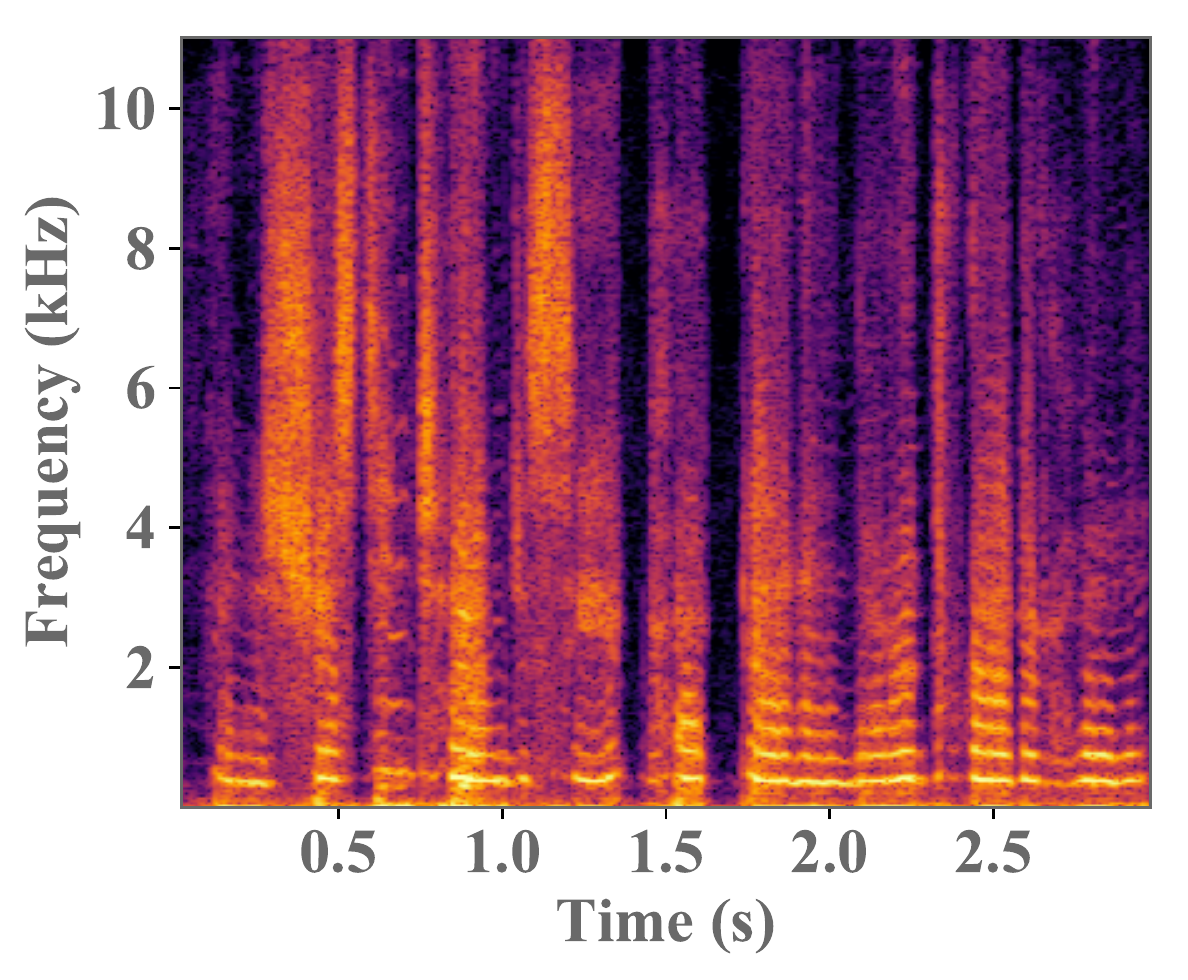} \\ \bottomrule
\end{tabularx}%
}
\caption{Comparison of intermediate cumulative outputs.}
\label{inter_output_1}
\end{figure*}

\subsection{Deformable Dilated Convolution}
The idea of a dynamically updated attention mechanism, which makes a sequential network know ``where to look'' at each time step, is not new. Generative models~\cite{gsrnn, ntm, draw} that utilize differentiable attention mechanisms to constrain the read and write operations of the network to specific parts of the scene have been proposed. To handle the limitation of the fixed geometric structure of the CNNs, the authors of~\cite{stn} proposed a learnable spatial transformation of the input feature maps of the CNNs to regularize the input of each CNN layer. Moreover, the authors of~\cite{dcn} proposed a deformable convolution to enable the freeform deformation of the CNN sampling grid. The deformable convolution gives the network an adaptive {\it receptive filed} that focuses on different locations of the input feature map corresponding to the current conditions.

Since the offsets of the grid sampling locations in PDCNN are derived from the $F_{0}$ values, the proposed PDCNN is a special case of a deformable CNN. As a deformable CNN with few additional parameters and computations, the PDCNN is implemented with a simple indexing technique\footnote{https://github.com/bigpon/QPPWG} without a large extra computational cost. As shown in Table~\ref{tb:rtf}, the average real-time factor (RTF) of the QPPWG\_20 inferences is similar to that of PWG\_20 and less than that of PWG\_30 when running on an Intel Xeon Gold 6142 CPU (2.60~GHz and 32 threads). However, because of the different indexing processes of each CNN kernel, the parallelization of the CNN computation on a GPU is degraded. As shown in Table~\ref{tb:rtf}, although the model size of QPPWG\_20 is only 70~\% of that of PWG\_30, the QPPWG\_20 model has 170~\% of the training time and 130~\% of the inference time of the PWG\_30 model when using an Nvidia TITAN V GPU. However, since the RTF of the PWG generation is much less than one, the additional inference time of QPPWG is insignificant.

\begin{figure}[t]
\fontsize{8pt}{9pt}
\selectfont
{%
\begin{tabularx}{1.0\columnwidth}{@{}XX@{}}
\toprule
  \centering\arraybackslash{QPPWG$af$\_20} 
&  \centering\arraybackslash{QPPWG${fa}$\_20} \\ \midrule
(a) $1/2\times F_0$ & (d) $1/2\times F_0$ \\ 
  \includegraphics[width=0.45\columnwidth]{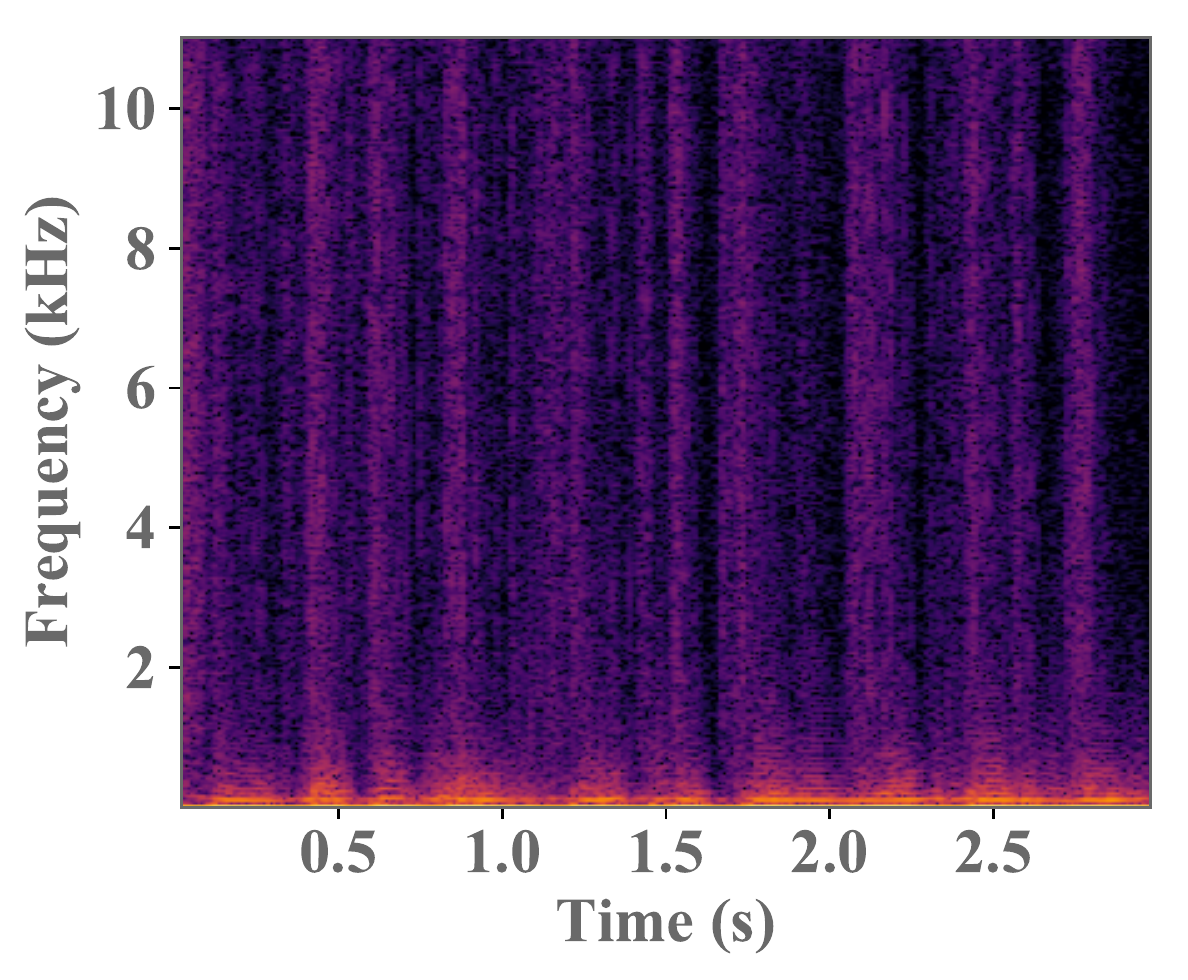}
&  \includegraphics[width=0.45\columnwidth]{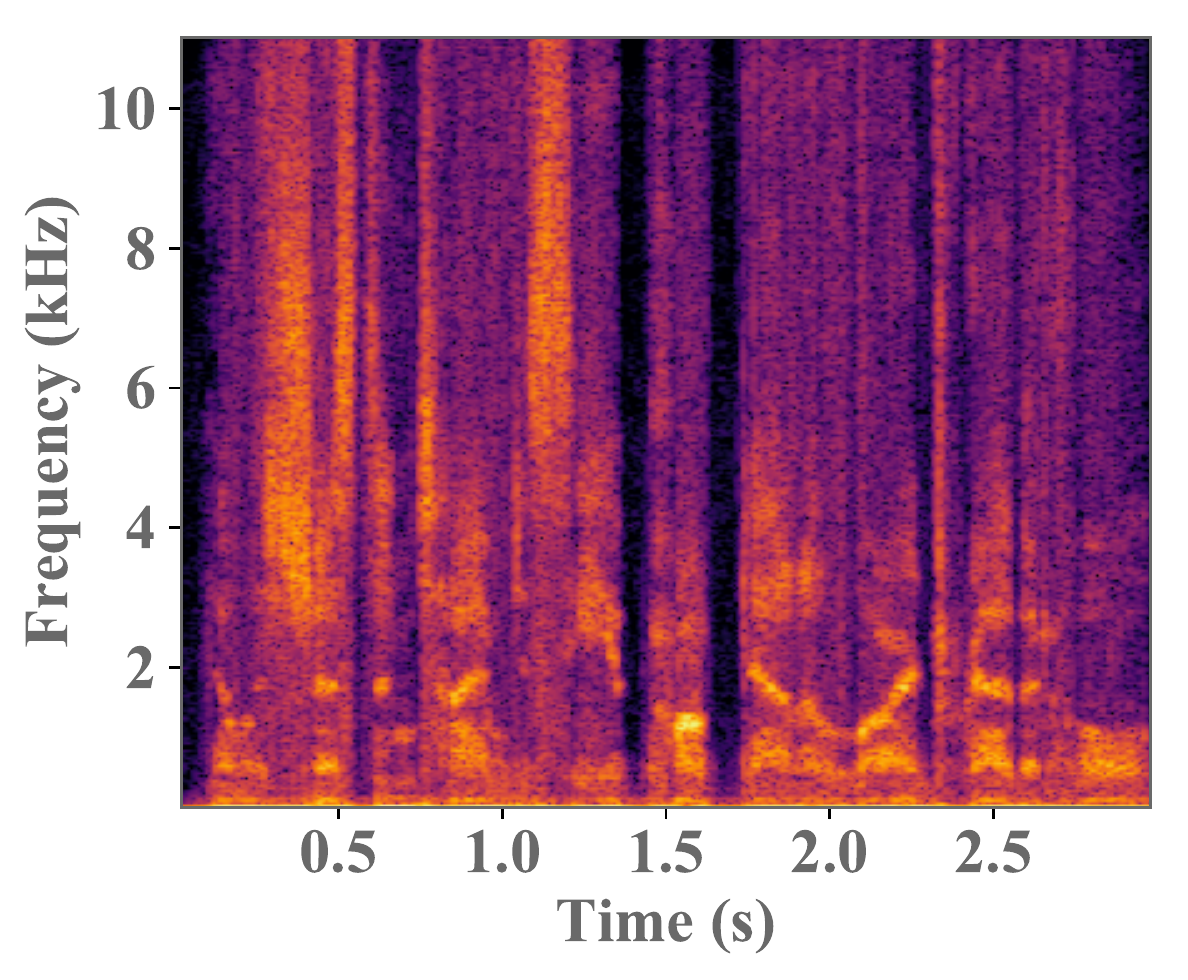} \\
(b) $1\times F_0$ & (e) $1\times F_0$ \\ 
  \includegraphics[width=0.45\columnwidth]{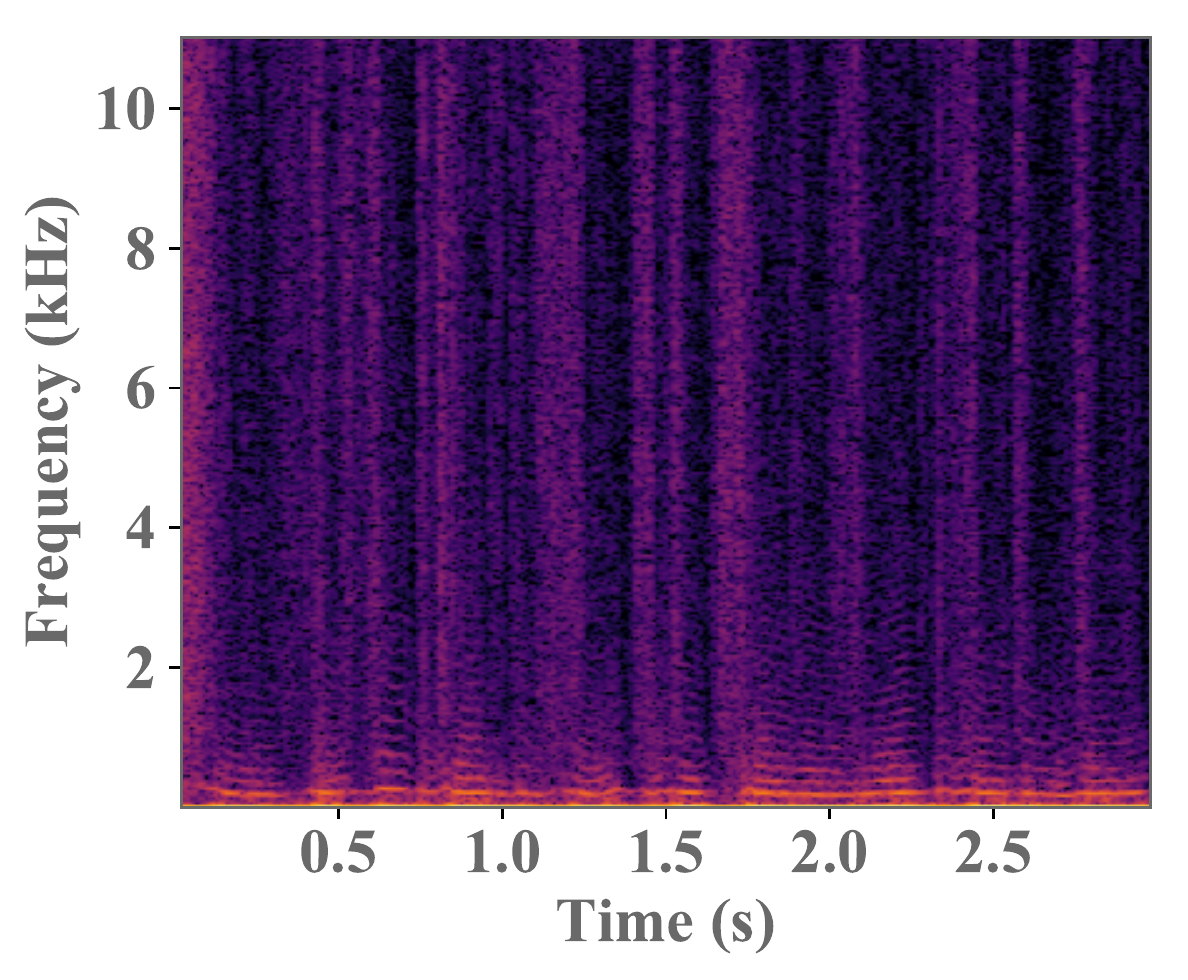}
& \includegraphics[width=0.45\columnwidth]{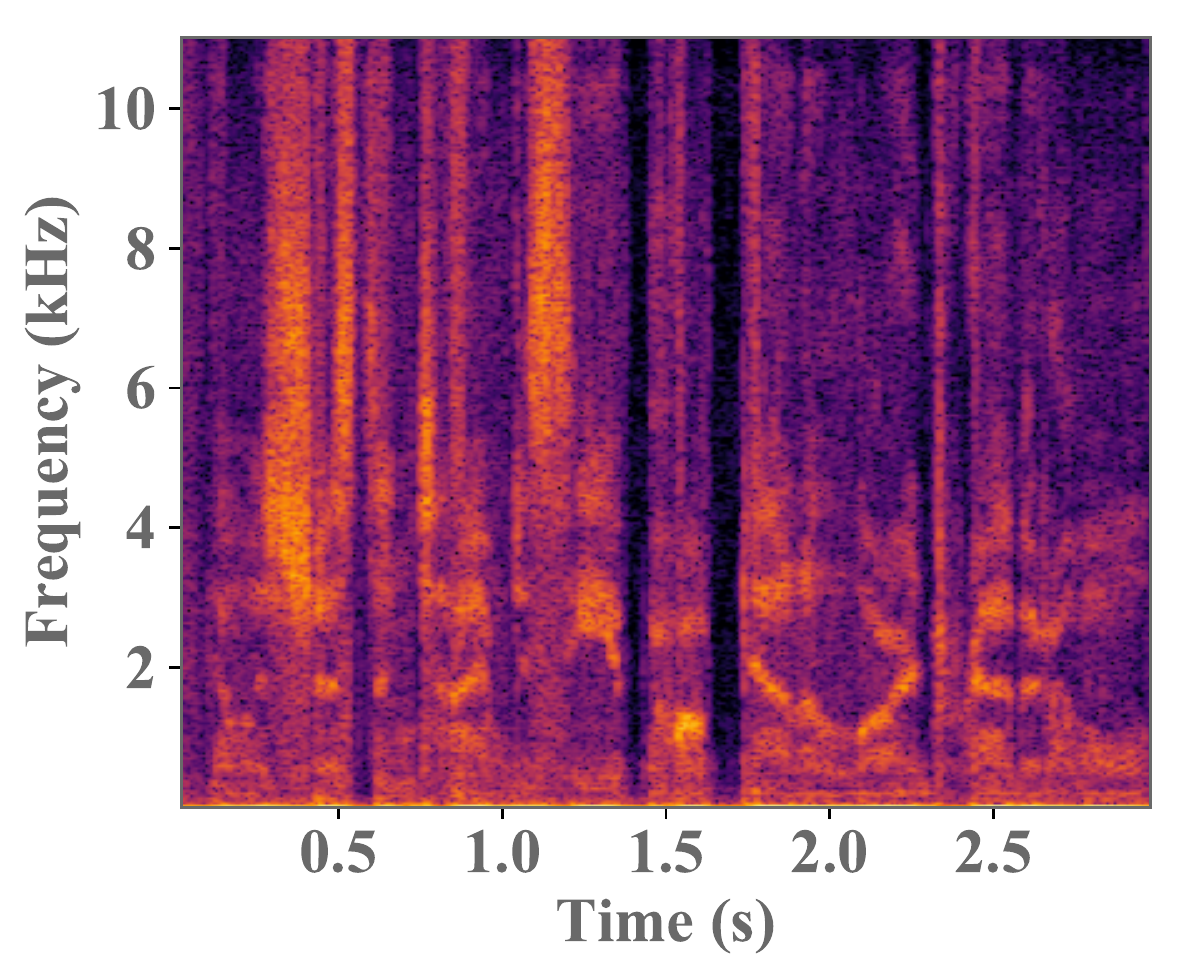} \\
(c) $2\times F_0$ & (f) $2\times F_0$ \\ 
  \includegraphics[width=0.45\columnwidth]{figs/qpaf_9_f0_2_00.pdf}
& \includegraphics[width=0.45\columnwidth]{figs/qpfa_9_f0_2_00.pdf} \\
\bottomrule
\end{tabularx}%
}
\caption{Comparison of intermediate cumulative outputs of 1--10 blocks with different $F_{0}$ scaled ratios.}
\label{inter_output_2}
\end{figure}

\subsection{Understanding of QP Structure}
Because of the direct waveform outputs of PWG/QPPWG, we can easily dissect the models to explore the internal speech modeling mechanisms. Specifically, the raw waveform outputs of the PWG/QPPWG models are the cumulative results of the skip connection outputs from the residual blocks. Therefore, the speech modeling behavior of the residual blocks can be explored via the visualized intermediate outputs of partial residual blocks. Spectrograms of the intermediate outputs of the cumulative residual blocks are presented in Fig.~\ref{inter_output_1}. For the PWG vocoder results (Figs.~\ref{inter_output_1} (a)--(d)), the spectrogram contains more details and textures as the number of cumulative residual blocks increases. In contrast to the PWG vocoder, which gradually adds both harmonic and non-harmonic components to the spectrogram, the first 10 adaptive blocks of the QPPWG${af}$ vocoder mostly focus on modeling the harmonic components as shown in Fig.~\ref{inter_output_1} (f). By contrast, the first ten fixed blocks of the QPPWG${fa}$ vocoder mostly generate the non-harmonic part of the speech as shown in Fig.~\ref{inter_output_1} (j). The results confirm our assumption that the adaptive blocks with the PDCNNs primarily model the pitch-related speech components with long-term correlations, while the fixed blocks with the DCNNs mainly focus on the spectrum-related components with short-term correlations.

In addition, to explore the behaviors of the adaptive and fixed blocks for different scaled $F_{0}$ features, comparisons among the visualized cumulative outputs of the first 10 residual blocks from the QPPWG${af}$ and QPPWG${fa}$ vocoders are presented. The spectrograms of QPPWG${af}$ shown in Figs.~\ref{inter_output_2} (a)--(c) have similar structures along the time axis but increasingly stretched harmonic structures along the frequency axis as $F_{0}$ increases. By contrast, despite the different $F_{0}$ scaled ratios, both the frequency and temporal structures of the spectrograms of QPPWG${fa}$ shown in Figs.~\ref{inter_output_2} (d)--(f) are similar. The results imply that the adaptive blocks primarily model the pitch-dependent harmonic components and the fixed blocks mainly focus on the pitch-independent non-harmonic components. Furthermore, although the QPPWG vocoder is a unified NN-based waveform generative model, the generative mechanism of its QP structure is similar to that of a source-filter model. The cascaded adaptive (pitch-dependent) and fixed macroblocks of the QP structure are analogous to the excitation generation and spectral filtering of the source-filter model. In conclusion, because a vocoder is assumed to have the capability for independently controlling each speech component, the QPPWG vocoder is more consistent with the definition of a vocoder. The QPPWG vocoder with the QP structure also attains a more tractable and interpretable architecture. More details of the visualized intermediate outputs can be found on our demo page~\cite{demo}.

\section{Conclusion} \label{conclusion}
To improve the pitch controllability of the PWG vocoder, we propose a QPPWG vocoder to introduce the prior pitch information to the network using the QP structure. Using the proposed non-AR PDCNN, the network architecture is dynamically adapted to the input $F_{0}$ feature of each input sample. Both objective and subjective experimental results show the effectiveness of the QP structure for the PWG vocoder. The QPPWG vocoder outperforms the PWG vocoder in pitch accuracy and speech quality for unseen scaled $F_{0}$ features while attaining a comparable speech quality to the PWG vocoder for natural $F_{0}$ features. Because of the more efficient {\it receptive field} expansion by PDCNNs, the model size of the QPPWG vocoder is only 70~\% of that of the PWG vocoder. Moreover, the visualized intermediate outputs of QPPWG vocoders confirm our assumption that adaptive blocks mainly model long-term correlations and fixed blocks focus on short-term correlations. To summarize, the proposed QPPWG vocoder is a fast and simple waveform generative model with higher pitch controllability, smaller model size, and better interpretability and tractability than vanilla PWG. The effectiveness of the QPPWG vocoder also indicates the generality of the QP structure for different CNN-based speech generative models.


\bibliography{mybib}
\bibliographystyle{IEEEtran}

%

\begin{IEEEbiography}
[{\includegraphics[width=1in,height=1.25in,clip,keepaspectratio]{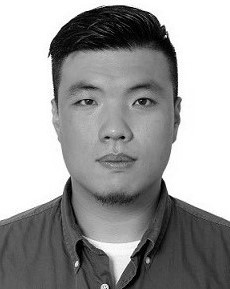}}]{Yi-Chiao Wu}
received his B.S and M.S degrees in engineering from the School of Communication Engineering of National Chiao Tung University in 2009 and 2011, respectively. He worked at Realtek, ASUS, and Academia Sinica for 5 years. Currently, he is pursuing his Ph.D. degree at the Graduate School of Informatics, Nagoya University. His research topics focus on speech generation applications based on machine learning methods, such as voice conversion and speech enhancement.
\end{IEEEbiography}

\begin{IEEEbiography}
[{\includegraphics[width=1in,height=1.25in,clip,keepaspectratio]{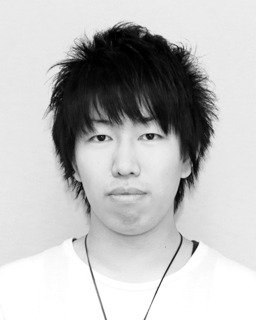}}]{Tomoki Hayashi}
received the B.E. degree in engineering and the M.E. and Ph.D. degrees in information science from Nagoya University, Japan, in 2014, 2016, and 2019, respectively. His research interests include statistical speech and audio signal processing. He is currently working as a postdoctoral researcher at Nagoya University and the chief operating officer of Human Dataware Lab. Co., Ltd. He received the IEEE SPS Japan 2020 Young Author Best Paper Award.
\end{IEEEbiography}


\begin{IEEEbiography}
[{\includegraphics[width=1in,height=1.25in,clip,keepaspectratio]{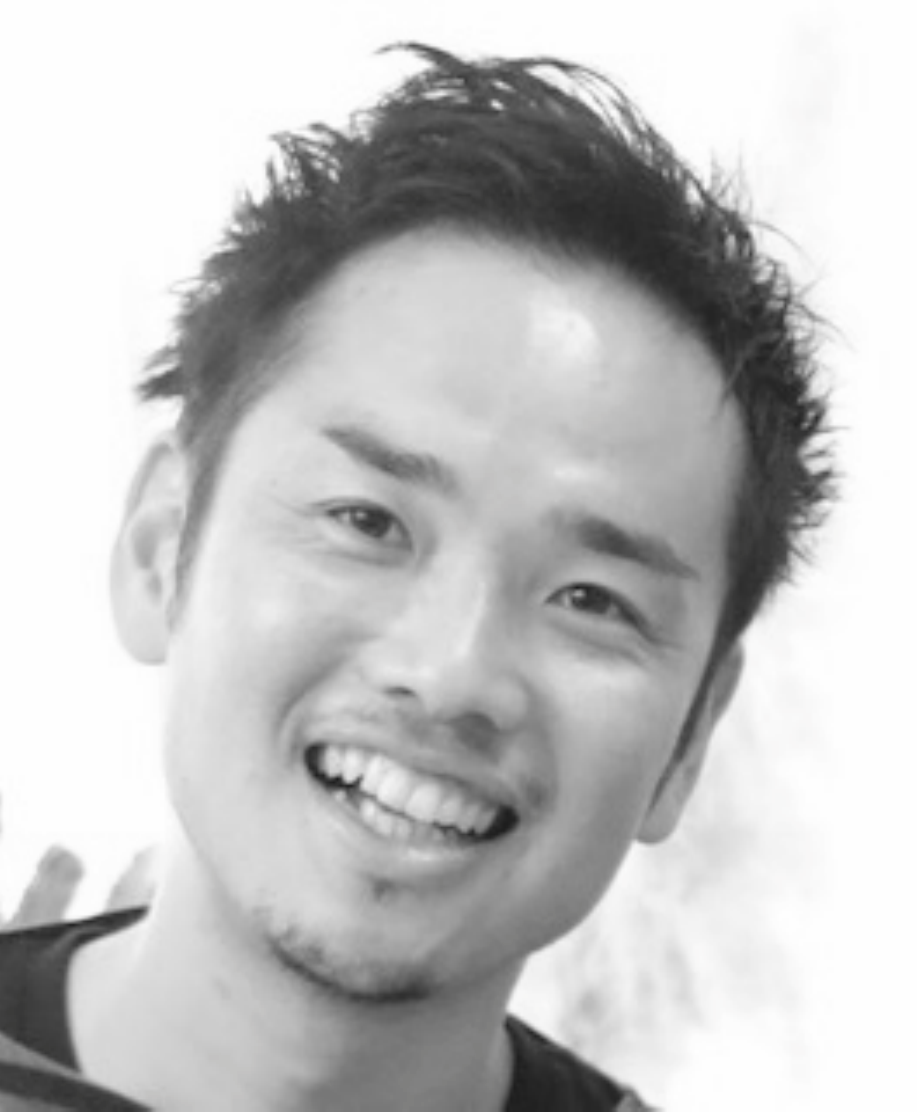}}]{Takuma Okamoto}
received the B.E., the M.S., and the Ph.D. degrees from Tohoku University, Japan, in 2004, 2006,
and 2009, respectively. From 2009, he was a postdoctoral research fellow at Tohoku University, Japan. During 2012 to 2020, he was a researcher at the National Institute of Information and Communications Technology, Japan, and he is currently a senior researcher there. His main research fields are sound field synthesis based on acoustic signal processing and speech synthesis based on neural networks. He
received the 32nd Awaya Prize Young Researcher Award and the 57th Sato Prize Paper Award from the Acoustical Society of Japan (ASJ) in 2012 and 2017, respectively. He is a member of ASJ.
\end{IEEEbiography}

\begin{IEEEbiography}
[{\includegraphics[width=1in,height=1.25in,clip,keepaspectratio]{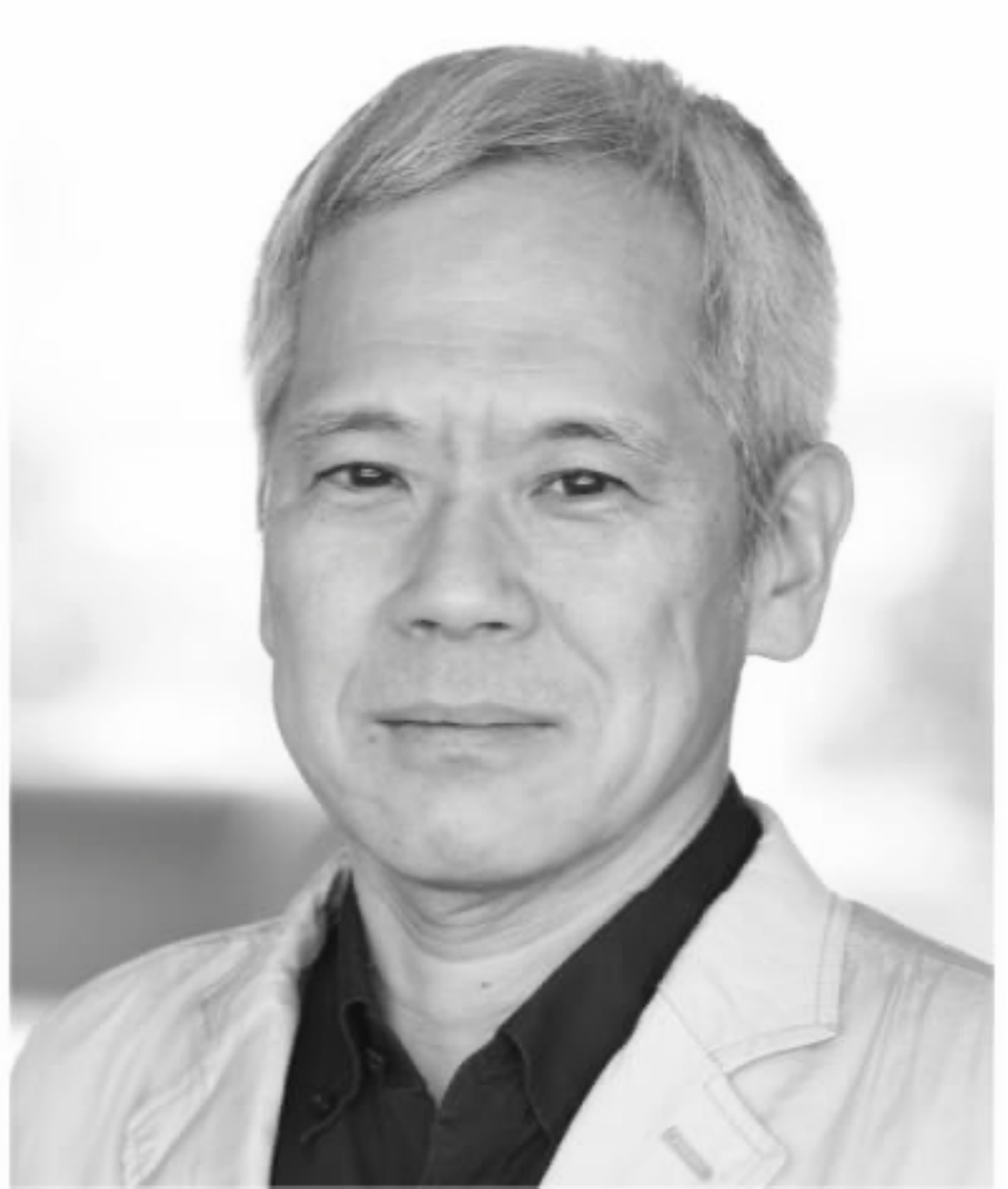}}]{Hisashi Kawai}
received the B.E., M.E., and D. E. degrees in electronic engineering from The University of Tokyo, Tokyo, Japan, in 1984, 1986, and 1989, respectively. He joined the Kokusai Denshin Denwa Co. Ltd., Tokyo, Japan, in 1989. He was with the ATR Spoken Language Translation Research Laboratories, Brisbane, QLD, Australia, from 2000 to 2004, where he was engaged in the development of text-to-speech synthesis system. From October 2004 to March 2009 and from April 2012 to September 2014, he was with the KDDI R\&D Laboratories, Fujimino, Japan, where he was engaged in the research and development of speech information processing, speech quality control for telephone, speech signal processing, acoustic signal processing, and communication robots. From April 2009 to March 2012 and since October 2014, he has been with the National Institute of Information and Communications Technology, Tokyo, Japan, where he is engaged in the development of speech technology for spoken language translation. He is a member of the Institute of Electronics, Information and Communication Engineers, Tokyo, Japan, the Acoustical Society of Japan, Tokyo, Japan, and the Institute of Electrical and Electronics Engineers, Piscataway, NJ, USA. (Based on document published on 5 April 2018).
\end{IEEEbiography}

\begin{IEEEbiography}
[{\includegraphics[width=1in,height=1.25in,clip,keepaspectratio]{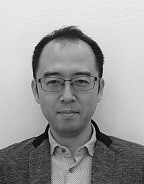}}]{Tomoki Toda}
is a Professor of the Information Technology Center at
Nagoya University, Japan. He received the B.E. degree from Nagoya
University in 1999, and the D.E. degree from the Nara Institute of
Science and Technology (NAIST), Japan, in 2003. He was a Research Fellow
of the Japan Society for the Promotion of Science from 2003 to 2005. He
was then an Assistant Professor (2005–2011) and an Associate Professor
(2011–2015) at NAIST. His research interests include statistical
approaches to speech, music, and environmental sound processing. He
received the IEEE SPS 2009 Young Author Best Paper Award and the 2013
EURASIP-ISCA Best Paper Award (Speech Communication Journal).
\end{IEEEbiography}




\end{document}